\documentclass{aa}  

\usepackage{subcaption}
\usepackage{caption}
\usepackage{tablefootnote}
\usepackage{graphicx}
\usepackage{txfonts}
\usepackage{xcolor}
\usepackage{colortbl}
\usepackage{tabularx}
\usepackage{appendix,natbib}
\usepackage{amsmath}
\usepackage{comment}
\usepackage[breaklinks=true]{hyperref} 
\hypersetup{colorlinks=true,citecolor=blue,linkcolor=blue,urlcolor=blue}

\newcommand{\sfgs}{star-forming galaxies}

\newcommand{\hz}{high-redshift }
\def \RCs{rotation curves}

\def \z1{$z\sim 1 $}

\newcommand\setrow[1]{\gdef\rowmac{#1}#1\ignorespaces}
\newcommand\clearrow{\global\let\rowmac\relax}
\clearrow

\def\apjl{ApJL}

\def\aap{A\&A}

\def\apjs{ApJS}

\begin{document}

\title{Tully-Fisher Relation of Late-type Galaxies at $0.6 \leq z \leq 2.5$}


\author{Gauri Sharma, 
    \inst{1,2,3,4,5,6} \fnmsep\thanks{Contact: gsharma@uwc.ac.za}
    \and 
    Varenya Upadhyaya, \inst{7}
    \and 
    Paolo Salucci \inst{4,5,6}
    \and
    Shantanu Desai \inst{7}
}

\institute{Observatoire Astronomique de Strasbourg, Université de Strasbourg, CNRS UMR 7550, F-67000 Strasbourg, France
    \and
    University of Strasbourg Institute for Advanced Study, 5 allée du Général Rouvillois, F-67083 Strasbourg, France
    \and
    Department of Physics and Astronomy, University of the Western Cape, Cape Town 7535, South Africa
    \and
    INFN-Sezione di Trieste, via Valerio 2, I-34127 Trieste, Italy
    \and
    IFPU Institute for Fundamental Physics of the Universe, Via Beirut, 2, 34151 Trieste, Italy
    \and 
    SISSA International School for Advanced Studies, Via Bonomea 265, I-34136 Trieste, Italy
    \and
    Department of Physics, Indian Institute of Technology, Hyderabad, Telangana-502284, India
}

\date{Received XXXX; accepted XXXX}

 
  \abstract
    {We present a study of the stellar and baryonic Tully-Fisher relation within the redshift range of $0.6 \leq z \leq 2.5$ utilizing observations of \sfgs. This dataset, as explored in \citet{GS23}, comprises of disk-like galaxies spanning a stellar mass range of $8.89 \leq \log(M_{star} \  [\mathrm{M_\odot}]) \leq 11.5$, baryonic mass range of $9.0 \leq \log(M_{bar}   [\mathrm{M_\odot}]) \leq 11.5$, and circular velocity range of $1.65 \leq \log(V_c \ [{\rm km/s}]) \leq 2.85$. Stellar masses of these objects are estimated using spectral energy distribution fitting techniques, while gas masses are determined via scaling relations. Circular velocities are directly derived from the Rotation Curves (RCs), after meticulously correcting for beam smearing and pressure support. Our analysis confirms that our sample adheres to the fundamental mass-size relations of galaxies and reflects the evolution of velocity dispersion in galaxies, in line with previous findings. This reaffirms the reliability of our photometric and kinematic parameters (i.e., $M_{star}$ and $V_c$), thereby enabling a comprehensive examination of the Tully-Fisher relation. To attain robust results, we employed a novel orthogonal likelihood fitting technique designed to minimize intrinsic scatter around the best-fit line, as required at \hz. For the STFR, we obtained a slope of $\alpha=3.03\pm 0.25$, an offset of $\beta = 3.34\pm 0.53$, and an intrinsic scatter of $\zeta_{int}=0.08$ dex. Correspondingly, the BTFR yielded $\alpha=3.21\pm 0.28$, $\beta=3.16\pm 0.61$, and $\zeta_{int}=0.09$ dex. Our findings suggest a subtle deviation in the stellar and baryonic Tully-Fisher relation with respect to local studies, which is most-likely due to the evolutionary processes governing disk formation.

    
    }
    


   \maketitle
\section{Introduction}\label{sec:c6-intro}
Scaling relations in galaxies refer to the empirical correlations between diverse observable properties, such as luminosity, mass, size, and rotational velocity. These relations offer invaluable insights into the fundamental physics and evolutionary dynamics shaping galaxies, and serve as rigorous benchmarks against which theoretical models of galaxy formation and evolution are tested. Among the various scaling relations, the Tully-Fisher relation holds a place of particular significance in galaxy evolution and cosmology. This  correlation acts as an analytical cornerstone, unraveling the complexities of galaxy dynamics and morphology, thereby deepening our understanding of interplay between physical properties of galaxies. 

In the realm of galaxy dynamics, the Tully-Fisher Relation (TFR) is one of the most studied empirical scaling relations that correlates the properties of luminous matter with those of the dark halo. In the traditional TFR, which originated from the seminal work of \citet[in][]{TFR1977}, the luminosity of galaxies scales with their characteristic velocity (i.e., circular velocity $V_c$) via a power-law, $L\propto \beta V_c^\alpha$, where $\alpha$ is the slope, and $\beta$ is the intercept in the relation. The slope indicates the extent of the circular velocity's dependency on the luminosity, while the quantity $\beta/\alpha$ represents the zero-point, which indicates the origin of the relation. In the local Universe, this relation is remarkably tight ($\alpha\sim 4$, $\beta/\alpha\sim 2$, $\sigma_{int} \lesssim 0.1$ dex) for star-forming disk galaxies \citep{TFR1977, Feast1994, belldejong2001, Karachentsev2002, Pizagno2007, Toribio2011, McGaugh2000, Sorce2013}. Consequently, it is widely used in redshift-independent distance measurements \citep{Giovanelli1997a, Ferrarese2000, Freedman2011, Sorce2013, Neill2014}, for example, knowing the luminosity and flux ($L$ and $F$), one can relate the observed flux to the distance ($D$) of the object via  $ F\propto L/4\pi D^2$. Furthermore, the TFR has played a significant role in determining cosmological parameters, particularly by enabling the measurement of the Hubble constant $H_0$ out to the local Universe  \citep{Giovanelli1997b, Tully2000, Masters2006}.

The TFR serves not only as a distance indicator in cosmology but also as a powerful tool for understanding the complex interaction between dark and luminous matter in galaxies. This is substantiated by a diverse range of observations \citep{Mathewson1992, McGaugh2000, McGaugh2005, Papastergis2016, Lapi2018} and simulations \citep{MoMao2000, Steinmetz1999}. The underlying rationale lies in the relationship between the circular velocity and the total gravitational potential of a galaxy, coupled with the luminosity serving as a tracer for the total stellar mass \citep{blumenthal1984, Mao1998, girardi2002}. An interaction between these physical quantities manifests as a correlation, thus giving rise to the well-known TFR. This foundational concept has also led to a generalized form of the TFR, expressed as $M\propto \beta V_c^\alpha$, where $M$ represents the galaxy's stellar or baryonic mass. This generalized TFR has undergone extensive study and exhibits remarkable tightness, particularly in the local Universe \citep{Verheijen2001, McGaugh2005, Blok2008, Stark2009, Foreman2012, Lelli2016, Papastergis2016, Lapi2018, Lelli2019}.

A note of caution is warranted when discussing the generalized TFR. In optical and infrared astronomy, luminosity primarily serves as a proxy for stellar mass, giving rise to the Stellar Tully-Fisher Relation (STFR). Conversely, at radio wavelengths, luminosity predominantly traces the mass of neutral hydrogen gas. When combined with the stellar mass, this provides an approximation of the total baryonic mass of a galaxy, $ M_{\text{bar}} \propto M_{\text{star}} + M_{\text{gas}}$, thereby leading to the Baryonic Tully-Fisher Relation (BTFR). It is noteworthy that the slope of the BTFR closely resembles that of the seminal TFR, with a typical value around 4 and an intrinsic scatter below 0.1 dex \citep[e.g.,][]{Lelli2019}. In contrast, the slope of the STFR generally ranges between 3 and 3.5-- depending on the wavelength range, accompanied by a larger intrinsic scatter of approximately \( \sim 0.25 \) dex \citep[e.g.,][]{Lapi2018}.

Some of the previous studies have investigated the seminal and generalized TFR of star-forming galaxies (SFGs) in cluster environments \citep{Ziegler2002, Bohm2004, Starkenburg2006}. These studies suggest that the slope of the TFR at intermediate redshifts ($z\sim 0.5$) is shallower compared to local measurements, which has prompted discussions on potential selection bias effects~\citep{Avila-Reese08,Gurovich10,Williams10,Mercier22,Catinella23}. However, other studies have reported little to no evolution in the seminal TFR slope from $z\sim 1$ to $z\approx 0$ \citep{Conselice2005, Kassin2007, Puech10, Miller2011,TorresFlores11, Zaritsky2014-dm,Abril21,Vergani12,McGaugh15}. Other high-$z$ studies, utilizing state-of-the-art Integral Field Unit (IFU) observations of isolated SFGs \citep{Puech2008,  Gnerucci2011, AT2016, Ubler2017} have found mixed results. Note however that  most of these works have  mainly focussed on  the evolution of the TFR zeropoint compared to the local Universe values after assuming a fixed slope.~\citet{Puech2008} found that  the slope in $K$-band  TFR at $z \sim 0.6$ is consistent with the local value  after allowing the slope to vary.
 ~\citet{Gnerucci2011} found a large scatter in the TFR at $z \sim 3$ and consequently used  a fixed slope having the  value same as that of the local Universe.~\cite{AT2016} studied the $K$-band TFR at  $z \sim 1$. They fit the TFR using both a fixed slope (obtained from the local Universe value) as well as keeping it as  a free parameter. When the slope was kept as a free parameter, significant differences  were found compared to the local Universe value (cf. Table 3 of ~\citealt{AT2016}). \citet{Ubler2017} studied the stellar and baryonic TFR at redshifts $z\sim 0.9$ and $z \sim 2.3$ by using a fixed slope (fixed to the value in ~\citealt{Lelli2016}) and looking for  the variation of  zeropoint with redshift.

A whole bunch of studies  have also been carried out on the  stellar TFR~\citep{Kassin2007,Cresci2009,Puech10,Williams10,TorresFlores11,Vergani12,AT2016,Price16,H17,Pelliccia2017,Ubler2017,Abril21,Catinella23}. In addition to fitting for the stellar TFR slope, many  works have also assumed a constant slope and evaluated the scatter~\citep{Cresci2009,Price16,Ubler2017}. Searches for an abrupt transition in the TFR slope using low redshift data have also been carried, which reported null results~\citep{Krishak22}.

Here, it is important to note that the early IFU studies have inherent uncertainties, primarily due to their 1D or 2D kinematic modeling approaches \citep[as reported by][]{ETD15}. This is because telescopes equipped with IFUs can achieve only a spatial resolution of $0.5-1.0 \arcsec$, while a galaxy at $z \sim 1$ typically has an angular size ranging from $2\arcsec - 3\arcsec$. As a result, a finite beam size leads to smearing of the line emission across adjacent pixels. Consequently, the gradient in the velocity fields tends to become flattened, and the line emission begins to broaden, creating a degeneracy in the calculation of rotation velocity and velocity dispersion. This observational effect is referred to as `Beam smearing,' which affects the kinematic properties of galaxies by underestimating the rotation velocity and overestimating the velocity dispersion. Therefore, it is essential to model the kinematics in 3D space, taking into account the beam smearing on a per-spaxel basis.  Recent studies by \citet{ETD16, GS21a}, and \citet{GS23} have modeled the kinematics of high-$z$ galaxies in 3D space, and demonstrated significant improvements in overall kinematics, including 2D velocity maps and position-velocity diagrams (i.e., observed rotation curves). It is noteworthy that although some of the  other  high-$z$ TFR studies have accounted for beam-smearing in the forward-modelling approach, none of them  have have fitted for kinematics in full 3D space similar to \citet{ETD16} and \citet{GS23}.

Moreover, at high-$z$, the Inter-Stellar Medium (ISM) in galaxies is highly turbulent \citep{Burkert2010, SW2019}. This turbulence within the ISM generates a force, which  counteracts gravity in the galactic disk via a radial gradient, which in turn suppresses the rotation velocity of gas and stars. This phenomenon is commonly referred to as `Asymmetric Drift' for the stellar component and `Pressure Gradient'  for the gas component, as defined in \citet{GS21a}. While the latter effect is generally negligible in local rotation-dominated galaxies (i.e., late-type galaxies), it is significantly observed in local dwarf and early-type galaxies \citep{val2007, Read2016, Anne2008}. The highly turbulent conditions of the ISM and the dominance of gas at high-$z$ \citep{Turner2017, HLJ17, SW2019} makes their velocity dispersion variable and anisotropic across the galactic scales \citep{Kretschmer2020}. Consequently, the observed rotation velocity measurements are underestimated throughout the galactic radius, and one may even observe a decline in the shape of rotation curves at high-$z$ \citep{Genzel2017, GS21a}. 

We point out that among the aforementioned high-$z$ TFR studies, only \citet{Ubler2017} has accounted for pressure gradient corrections by assuming a  constant and isotropic velocity dispersion. However, recent studies  of \hz observations \citep{GS21a} and simulations \citep{Kretschmer2020} indicate that pressure support corrections under the assumption of constant and isotropic velocity dispersion can lead to an overestimation of  the circular velocities. This is particularly relevant for galaxies with low rotation-to-dispersion ratios $(v/\sigma < 1.5)$. Given these findings, there is a compelling need to re-examine the TFR at \hz, employing more precise kinematic measurements as recommended in \citet{ETD16, GS21a}, and \citet{Kretschmer2020}. 

This study aims to revisit and refine our understanding of the TFR at high-redshifts. Specifically, we utilize a large dataset recently analyzed by \citet{GS23}, which models the kinematics using 3D forward modeling and incorporates the pressure gradient while allowing for varying and non-isotropic velocity dispersion. The aim of this work is to investigate the cosmic-evolution of TFR in star-forming galaxies-- disk-like systems, within the redshift range of $0.6\leq z\leq 2.5$. We focus on disk-like systems since they form and evolve predominantly at $z\leq 1.5$ and exhibit homogeneous and controlled evolution \citep[e.g.,][]{Lagos2017}. Thus, these systems serve as a valuable tool to infer the cosmic evolution of baryons and dark matter. 
At $z\approx 1$, nearly 50\% of the Universe's stellar mass assembles in galactic halos \citep{PrezGonzlez2008}, and this marks the peak of cosmic star-formation density \citep[][and references therein]{Madau2014}. Therefore, it is crucial to compare the baryonic and dark matter properties of galaxies at $0.6\leq z\leq 2.5$ with those in the local Universe. This comparison provides insights into (1) the evolution of disk-like systems after their formation at $z\leq 1.5$ and (2) the nature of dark matter because these systems are more-or-less in dynamical equilibrium.

This paper is organized as follows, Section~\ref{sec:c6-data} discusses the dataset, relevant parameters of STFR and BTFR relations, and assess the quality of these parameters using fundamental scaling relations. In Section~\ref{sec:TFR}, we present the STFR and BTFR relations, and in Section~\ref{sec:c6-discussion} we discuss these relations in comparison with previous studies. Finally, in Section~\ref{sec:c6-summary} we summarize our work and conclude main findings. In this work, we have assumed a flat $\Lambda$CDM cosmology with $\Omega_{m,0} =0.27$, $\Omega_{\Lambda,0}=0.73$ and $H_0=70 \ km \ s^{-1} Mpc^{-1}$.

\section{Data}
\label{sec:c6-data}
In this study, we make use of the dataset recently examined by \citet[][hereafter GS23]{GS23}. As discussed in GS23, this sample was initially selected based on the assessment of kinematic modeling outputs. Briefly, kinematic modeling was based on the following primary criteria: (1) confirmed H$\alpha$ detection and spectroscopic redshift, (2) inclination angles within the range of $25^{\circ} \leq \theta_i \leq 75^{\circ}$, and (3) SNR $ > 3$ (in $H\alpha$ datacubes). GS23 employed the 3DBarolo code to model the kinematics, allowing for beam smearing corrections and inclination within a 3D space. This results in velocity maps, major and minor axis position-velocity (PV) diagrams, surface brightness curves, rotation curves, and velocity dispersion curves. Following the kinematic modeling outcomes GS23 implemented secondary selection criteria, according to which  galaxies were excluded if they met the following conditions: (1) 3DBarolo run did not succeed, (2) No mask was created, implying 3DBarolo's failure to mask the true emission due to a moderate signal-to-noise; (3) maximum observed radius smaller than the point spread function, i.e.,  $\mathrm{R_{\rm max}< PSF}$, indicating 3DBarolo's inability to create rings and hence fails to produce kinematic models; (4) $\mathrm{R_{\rm max}=PSF}$, in this case resulting kinematic models provide only two measurements in \RCs, which were insufficient for dynamical modeling or reliable measurements of circular velocities. This secondary selection criteria resulted in a final sample of 263 galaxies, comprising 169 from KROSS, 73 from KMOS3D, and 21 from KGES. For the distribution of relevant physical quantities of the final sample we refer the reader to \citet[][Fig.~4]{GS23}. 

\begin{figure}
	\begin{center}
		\includegraphics[angle=0,height=6.truecm,width=9.0truecm]{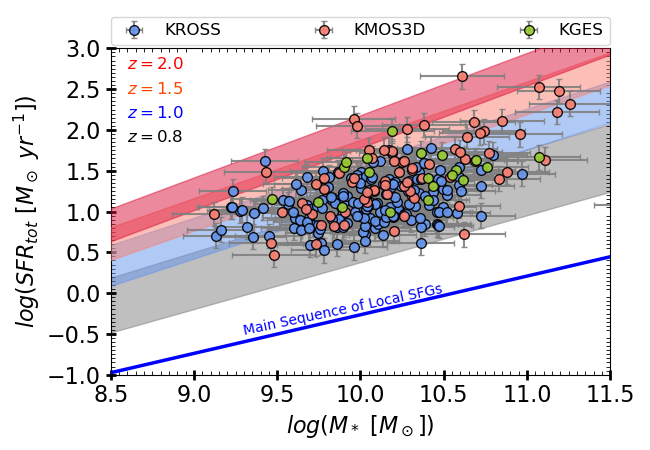}  
		\caption{Main sequence of local star-forming galaxies is shown by solid blue line, and for redshift 1.0, 1.5, and 2.0, shown by  blue, pink, and red shaded areas, respectively. The KMOS3D, KGES and KROSS data is shown by red, green, and blue filled circles. Hereafter, we refer this to full dataset as  GS23 and depict in blue color throughout the work.}
		\label{fig:Mstar-sfr}
	\end{center}
\end{figure}

The rotation curves inferred from 3DBarolo are further corrected for pressure support through the `Pressure Gradient Correction,' method as established by \citet{GS21a}, and referred  to as intrinsic rotation curves. We utilize these intrinsic rotation curves to estimate the circular velocities ($V_c$) of galaxies. The velocity dispersion ($\sigma$) is an average value estimated from velocity dispersion curves obtained from 3DBarolo, for more details we refer the reader to \citet{GS21a} and \citet{GS23}. GS23 sample spans a stellar mass range of $8.89 \leq \log (M_{\rm star} \ [\rm M_\odot]) \leq 11.5$, effective radii $-0.2 \leq \log(R_e \ [\rm kpc]) \leq 0.85$, star formation rates between $0.49 \leq \log\left(\mathrm{SFR}\ [\mathrm{M_{\odot}\ yr^{-1}}]\right) \leq 2.5$, and a redshift range of $0.6 \leq z < 2.5$. This sample is a fair representative of main-sequence star-forming galaxies, shown in Figure~\ref{fig:Mstar-sfr}. In the subsequent sections, we briefly examine the circular velocity and velocity dispersion estimates, discuss the baryonic mass estimates, and justify the accuracy of photometric and kinematic properties relevant for TFR study.

\begin{figure}
		\includegraphics[width=\columnwidth]{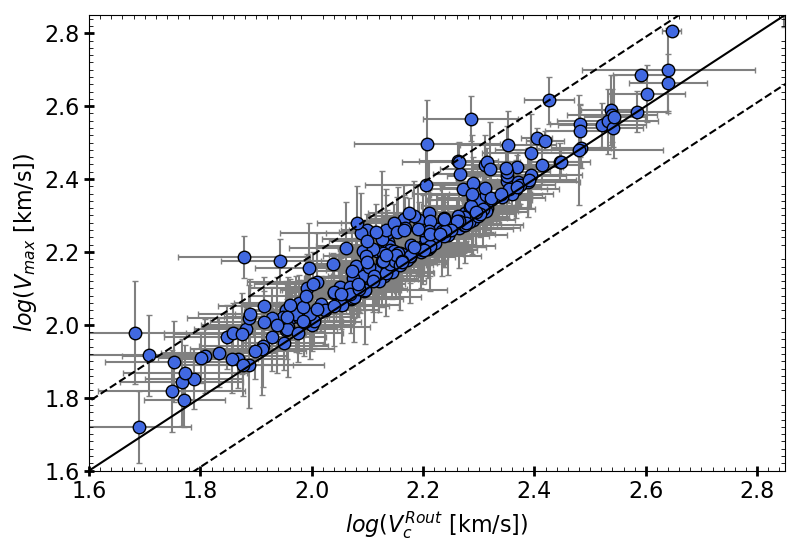}
		\caption{The comparison of circular velocities, $V^{Rout}_c$ and $V_{max}$. The black solid line shows the one-to-one relation followed by dashed lines showing the $1\sigma$ intrinsic scatter around this line. Since the measurements of $V^{Rout}_c$ and $V_{max}$ correlate within $1\sigma$, it suggests that both velocity measurements are good proxies for circular velocity of galaxies. In the analysis we refer to $V_c^{Rout} = V_c$ as the circular velocity of the object.}
		\label{fig:Vc-Vmax}
\end{figure}

\begin{figure}
		\includegraphics[width=\columnwidth]{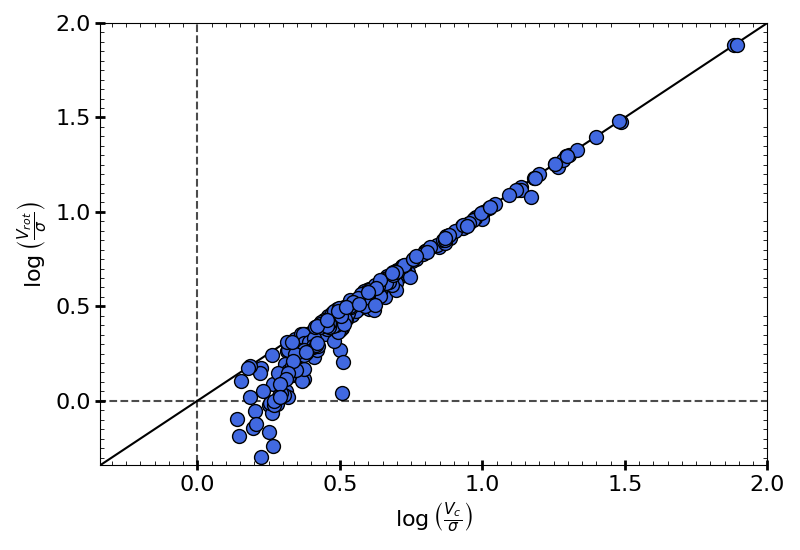}
		\caption{Intrinsic and pressure-support-corrected rotation-to-dispersion ratio ($V/\sigma$), plotted on the y- and x-axes, respectively. The solid black line represents the one-to-one relation between the two quantities. The vertical and horizontal dashed lines indicate the $V/\sigma > 1$ limit for intrinsic and pressure-support-corrected rotation-to-dispersion ratios, respectively. This figure indicates that, after pressure support corrections, none of the GS23 galaxies exhibit dispersion-dominated characteristics. Therefore, we utilize the entire GS23 sample for the TFR study.}
		\label{fig:VoverSigma}
\end{figure}

\subsection{Velocity Measurements} \label{sec:Vc}
In this study, we have examined the circular velocity of rotation curves at three distinct scale lengths, specifically $R_e$, $R_{\rm opt}$, and $R_{\rm out}$ ($\approx 5\ R_{\rm D}$), which we denote as $V^{Re}_c$, $V^{Ropt}_c$, and $V^{Rout}_c$, respectively.\footnote{For an exponential thin disk, the stellar-disk radius is defined as $R_D = 0.59 \ R_e$. Under the same assumption, the scale length that encloses 80\% of the stellar mass is referred to as the optical radius and defined as $R_{\rm opt}=3.2\ R_D$. For more comprehensive details, we refer the reader to \citet{PS1996}.} It is worth noting that the effective radius for the majority of our sample falls below the resolution limit, which is approximately $4.0$ kpc with a median seeing of $0.5\arcsec$. On the other hand, the optical radius remains on the verge of resolution limit. Thus, in order to be conservative, we only utilized circular velocity measurements that were obtained at $R_{out}$. This is one of the reasons of not plotting TFR for $V_c(2.2 R_D)$ as  adopted in previous studies \citep[e.g.,][]{Ubler2017, AT2019a}. However, the choice of $V_c(2.2 R_D)$ aims to capture the flat portion of the rotation curves, akin to $V_{c}^{Rout}$ in our case, which represents the circular velocity in the outer regions of the rotation curves assumed to be flat. Finally, it is important to remark that  $\sim 92\%$ and 65\% of galaxy rotation curves are sampled up to $R_{\rm opt}$ and $R_{\rm out}$, respectively. In cases where the rotation curve is not sampled up to the reference radius, we interpolate (or extrapolate) the velocity estimates. Our approach is as follows: (1) If $R_{opt}$ exceeds $R_{last}$ (the maximum observed radius), $V_{c}$ is computed at $R_{last}$. (2) If $R_{out} > R_{last}$, $V_{c}$ is computed at $R_{opt}$. This approach ensures that we remain within the outer regions of galaxies, which are assumed to have flat rotation curves based on local observations. Note that we did not interpolate (or extrapolate) the velocities beyond the maximum observed radius. Moreover, for interpolation, we do not employ any specific functional form of the rotation curve; instead, we utilize \texttt{numpy.interp} routine. This ensures that if the rotation curve is declining, it will continue to decline, and vice versa.


Additionally, it's worth noting that TFR studies in the local Universe occasionally utilize the maximum velocity of the system \citep[e.g.,][]{Lelli2019}. Therefore, we also examined the maximum velocity in relation of $V_c^{Rout}$. We extracted the maximum circular velocity ($V_{\rm max}$) from the rotation curves. We note to the reader that $V_{max}$ is not the asymptotic rotation velocity and hence involves no interpolation/extrapolation. The results of this comparison are presented in Figure~\ref{fig:Vc-Vmax}. Our analysis revealed a strong positive correlation of $\sim 97\%$ between $V_{\rm max}$ and $V^{Rout}_c$, with an intrinsic scatter of $0.15$ dex. Although we have observed that $\sim 30\%$ of the sample exhibits $V_{max}$ values 0.1 dex higher than $V^{Rout}_{c}$, we consider to use $V^{Rout}_{c}$ as the circular velocity. The rationale behind this decision is the uncertainty in capturing the entire flat part of the \RCs\  at~ \hz. As a result, $V_{\rm max}$ might not accurately represent the maximum velocity of the galaxy. Hence, it can not be compared neither with the locals nor with \hz\ studies. Therefore, to maintain uniformity across the sample, treat all galaxies consistently, and facilitate comparisons with previous \hz\ studies \citep[e.g.,][]{AT2019a, Ubler2017}, we have chosen to utilize $V^{Rout}_c$ as the circular velocity, hereafter denoted as $V_c$. 

In Figure~\ref{fig:VoverSigma}, we show the rotation-to-dispersion ratio of before and after pressure support corrected GS23 sample. The velocities before pressure support corrections are referred to as rotation velocity ($V_{\rm rot}$), while after pressure support corrections its circular velocity ($V_{\rm c}$) of the system. We notice that, prior to the implementation of pressure support corrections, there were only 9 dispersion dominated galaxies (3 KMOS$^{\rm 3D}$, 1  KGES, and 5 KROSS). However, after applying pressure support corrections, none of these galaxies have $V_{\rm c}/\sigma < 1$, as depicted in Figure~\ref{fig:VoverSigma}. Therefore, we do not exclude these galaxies from our analysis. Hence, the full GS23 sample is a good representative of rotation supported systems, which we employ to study TFR. We remark to the reader that underlying assumptions of GS23 consist of three key criteria: 
\begin{enumerate}
\item Galaxies should be located on or around the star-forming main sequence; 
\item They should exhibit a disk-like morphology in high-resolution images, with no nearby neighbors within 150 kpc; and
\item The ratio of circular velocity to velocity dispersion ($V_c/\sigma >1$). 
\end{enumerate}
These three assumptions enable them to select disk-like galaxies from high-z sample. Notably, our main findings in Section~\ref{sec:TFR} are consistent with those of local studies \citep[e.g.,][]{Lapi2018, Reyes2011} which generally select the star-forming galaxies with $V_{\rm c}/\sigma >1$. However, we notice that previous high-$z$ studies apply higher $V_{\rm rot}/\sigma$ cuts to investigate the TFR, which we briefly discuss in Appendix~\ref{app:New-HZ-comp}.

\begin{figure*}
		\includegraphics[width=\columnwidth]{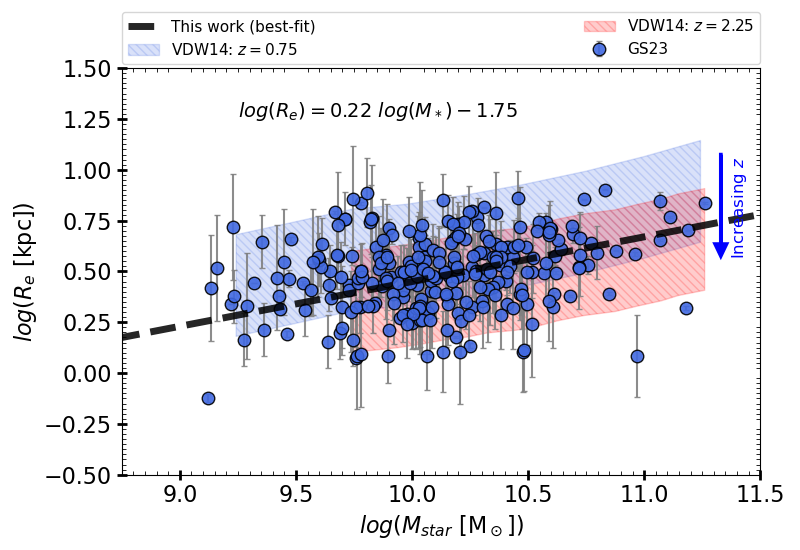}
  		\includegraphics[width=\columnwidth]{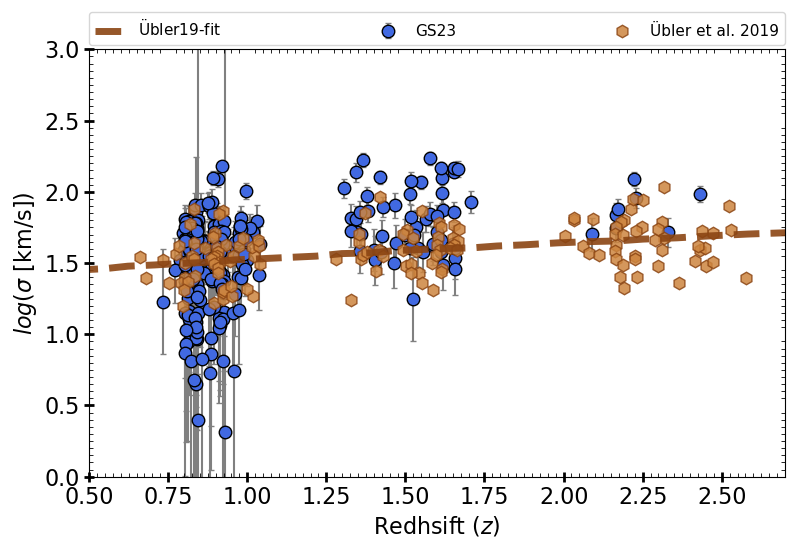}
		\caption{{\em Left panel:} Mass-size relation of late-type galaxies. The blue filled circles represents the data \citep[of][employed in this work]{GS23}, and black dashed line shows the best-fit (slope and offset printed on plot). The blue and red shaded areas represent the mass-size relation of \citet{Vanderwel2014} at $z=0.75$, and $z=2.25$, respectively. {\em Right panel:} Ionized gas velocity dispersion as a function of redshift. The blue filled circles and brown hexagons represent the \citet[][]{GS23} and \citet{Ubler2019} data, respectively. The brown dashed line represents the best fit of \citet{Ubler2019} work. We notice that 73 out of the 263 (27.7\%) galaxies from the \citet[][]{GS23} sample are common in the two datasets. }
		\label{fig:scaling-Relations}
\end{figure*}

\subsection{Baryonic Masses} \label{sec:Mbar}
Observations show that typical star-forming galaxies lie on a relatively tight, almost linear, redshift-dependent relation between their stellar mass and star formation rate, the so-called main sequence of star formation \citep[MS; e.g., ][]{Noeske2007, Whitaker2012, speagle14}. Most stars since $z\sim 2.5$ were formed on and around this MS \citep[e.g., ][]{Rodighiero2011}, and galaxies that constitute it, usually exhibit a rotating disk morphology \citep[e.g., ][]{ForsterSchreiber2006, Daddi2010b, Wuyts2011b}. Figure~\ref{fig:Mstar-sfr} shows the position of the GS23 sample on the main sequence of typical star-forming galaxies (MS), the analytical prescription for the centre of the MS as a function of redshift and stellar mass proposed in the compilation by \citet{speagle14}, as a function of stellar mass.  The figure shows that the all sources are on and around the main sequence between $0.65 \leq z \leq 2.5$. A normalized main sequence plot of this dataset is shown in \citet[][Fig.~3]{GS23}. This suggests that GS23 sample is a good representative of disk-like star-forming galaxies.

This enables us to estimate their molecular gas masses ($M_{H2}$) using the \citet{Tacconi2018} scaling relations, which provide a parameterization of the molecular gas mass as a function of redshift, stellar mass, and offset from the MS, stemming from a large sample of about 1400 sources on and around the MS in the range $z=0-4.5$ (cf. also \citealt{Genzel2015} and \citealt{Freundlich2019}). The scatter around these molecular gas scaling relations and the stellar mass induce a 0.3 dex uncertainty in the molecular gas mass estimates. The H2 mass of GS23 sample is $9.14 \leq \log(M_{\rm H2} \ [M_\odot]) \leq 10.63$, with an average molecular gas fraction of $f_{_{\rm H2}} = 0.12\pm 0.04$.
	
To calculate the atomic mass ($M_{HI}$) content of galaxies within the redshift range $0.6 \leq z \leq 1.04$, we use the HI scaling relation presented by \citet{Chowdhury2022}, which provides the first $M_{star}-M_{HI}$ relation at $z\approx 1$, encompassing 11,419 star-forming galaxies. The relation was derived using a stacking analysis across three stellar mass bins, each bin with a $4\sigma$ detection and an average uncertainty of $\sim 0.3$ dex. To compute the HI mass at $z> 1.04$, we employ the $M_{star}-M_{HI}$ scaling relation derived from a galaxy formation model under the $\Lambda \mathrm{CDM}$ framework \citep[for details see,][]{Lagos2011}. This scaling relation successfully reproduces both the HI mass functions \citep{Zwaan2005, Martin2010} and the $^{12}CO$ luminosity functions \citep{Boselli2002, Keres2003} at $z\approx 0$ with an uncertainty of around $0.25$ dex, as well as follows the observations of quasars  from $z=0-6.4$ (see Fig.12, \citealt{Lagos2011}). The HI Mass range of GS23 sample is $9.57 \leq \log(M_{\rm HI} \ [M_\odot]) \leq 11.05$, with an average atomic gas fraction $f_{_{\rm HI}} = 0.41\pm 0.13$. Finally, the total baryonic mass of galaxies is the sum of molecular and atomic gas : $M_{\rm bar} = M_{\rm H2}+1.33 M_{\rm HI}$, where the factor
of 1.33 accounts for the Helium content.



\begin{figure*}
		\includegraphics[width=\columnwidth]{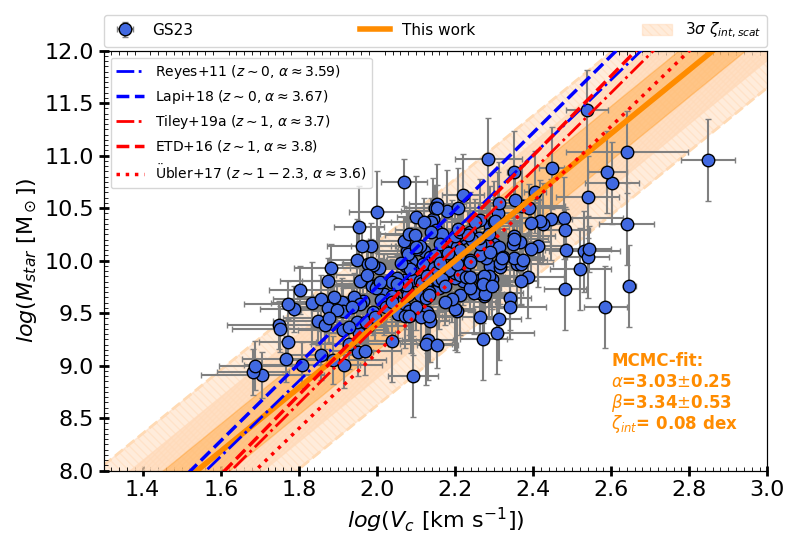}
		\includegraphics[width=\columnwidth]{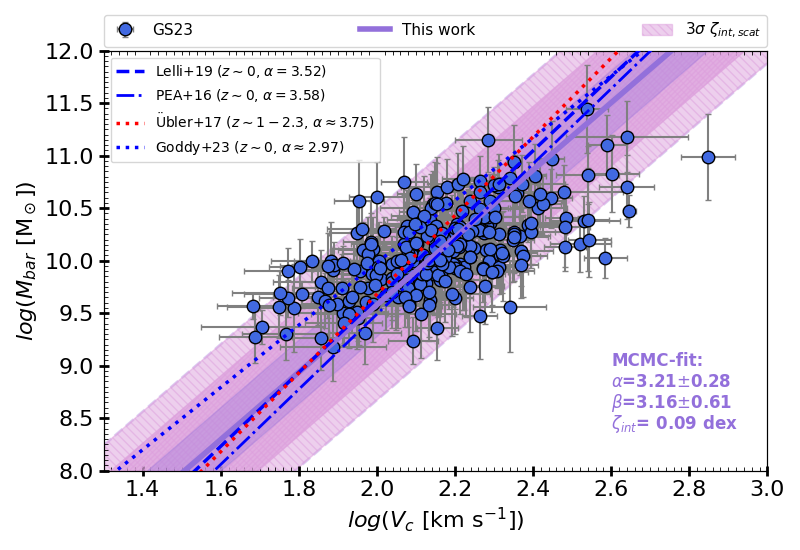}
		\includegraphics[width=\columnwidth]{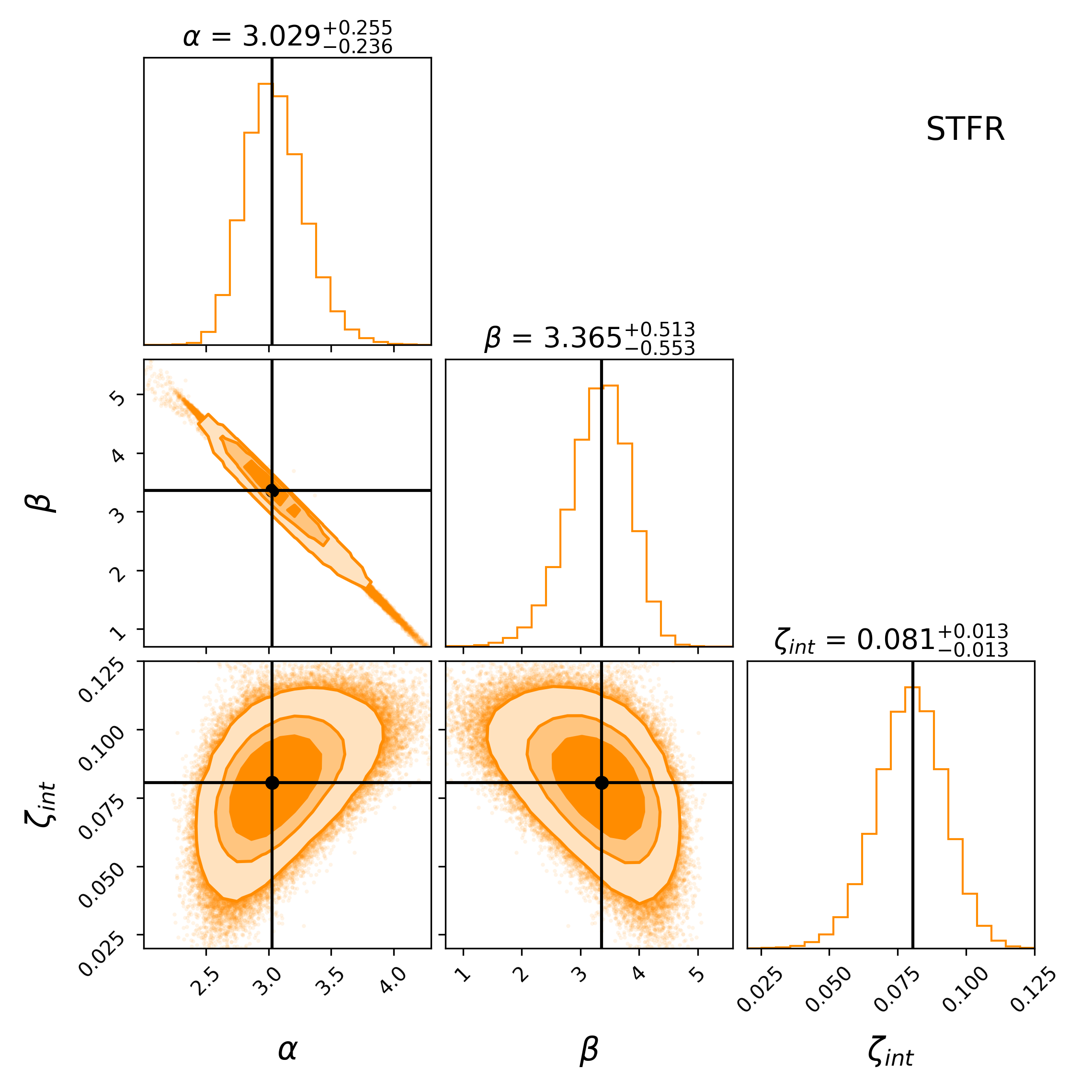}
		\includegraphics[width=\columnwidth]{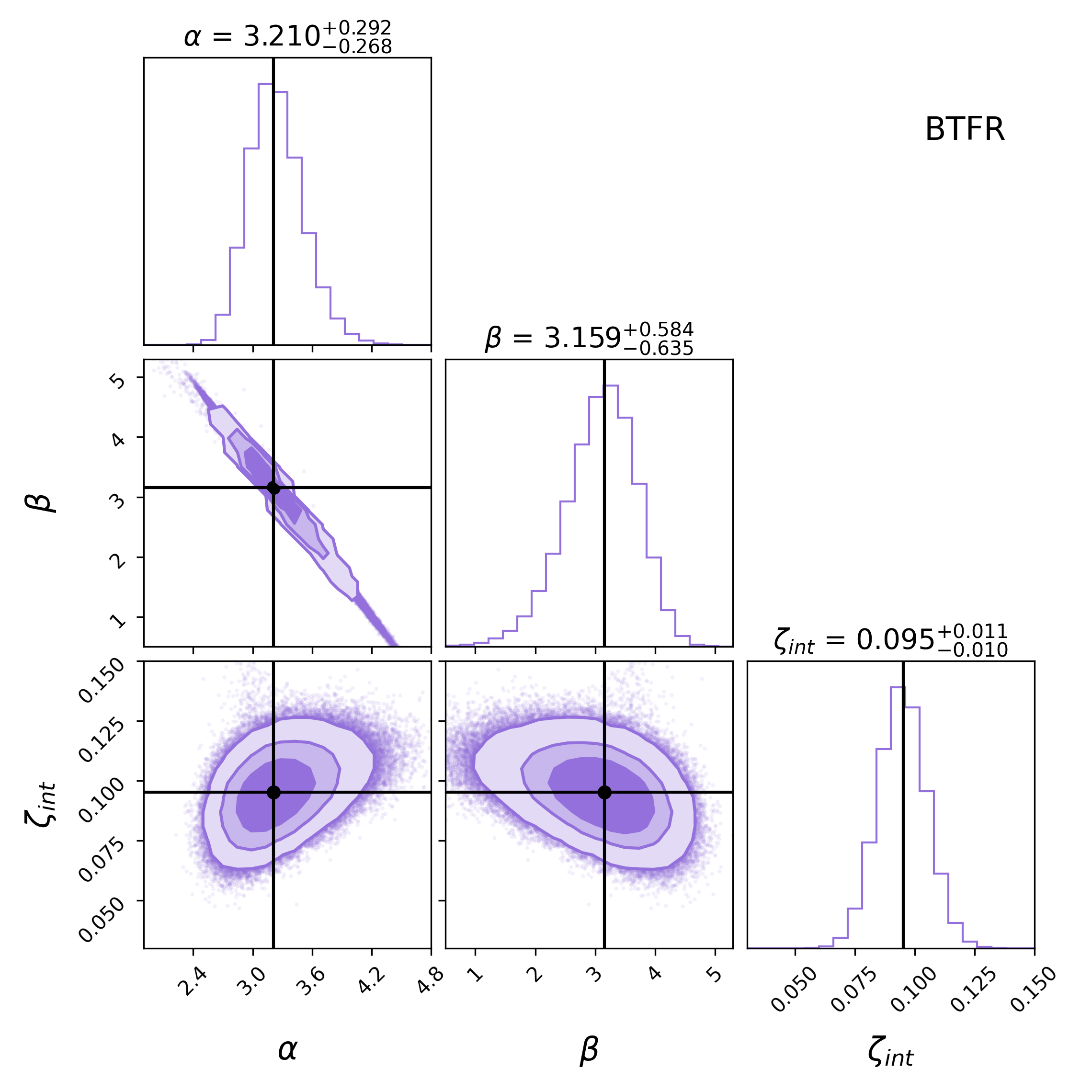}
		\caption{\textit{Upper Panel:} Stellar and Baryonic Tully-Fisher Relations (STFR and BTFR), presented in the left and right panels, respectively. The blue filled circles represent the data from \citet{GS23}, with gray error bars denoting uncertainties on each measurements. The solid orange and purple lines shows the best-fit curves obtained in this study using orthogonal likelihood, accompanied by the shaded regions representing the $3\sigma$ intrinsic scatter for the STFR and BTFR, respectively. The bottom right corner of each plot displays the best-fit parameters. Additionally, the blue lines correspond to comparisons with local studies, while red lines represent the high-redshift data, as indicated in the upper left legend of each plot. \textit{Lower Panel:} Posterior distributions (corner plots) resulting from the MCMC fitting process for the STFR and BTFR are shown in the left and right plots, respectively. The
contours within these corner plots illustrate the 68\%, 90\%, and 99\% credible intervals. For the reference of the reader, we also show vertical likelihood fits of STFR and BTFR in Figure~\ref{fig:TFR1_vert}, which shows a huge difference in the slope and zero-point of the relation with respect to orthogonal likelihood. We report a difference of about factor 2.}
		\label{fig:TFR1}
\end{figure*}

\subsection{Quality Assessment of Data}\label{sec:Qasses}
As shown in Sections~\ref{sec:Vc} and \ref{sec:Mbar}, the GS23 sample contains rotation supported systems, which lies on-and-around the main-sequence of star-forming galaxies. In this section, we focus on the quality assessment of our dataset, particularly emphasizing the verification of key scaling relations such as the mass-size relation and the redshift evolution of the velocity dispersion. The consistency of these relations serves as a benchmark for the overall integrity of our dataset and the subsequent analysis of TFR across cosmic time. 

{\bf Mass-Size Relation:}
In the local Universe, galaxies are broadly categorized into two main classes: early-type and late-type, commonly identified as the red-sequence and blue-cloud, respectively \citep{Gavazzi2010}. These classes exhibit distinct relationships between stellar-disc size and total stellar mass \citep{Shen2003}. However, for nearly a decade, cosmic evolution of the mass-size relation for galaxies was an open question, (e.g. early-type: \citealt[][]{Daddi2005, Vanderwel2008, Saracco2011, Carollo2013}; late-type: \citealt[][]{Mao1998, Barden2005, Mosleh2011}). Recently, with a large dataset of CANDELS survey, \citet{Vanderwel2014} statistically studied the mass-size relation of early- and late-type galaxies through the redshift range: $0<z<3$. Their findings indicate that while the intercept of the mass-size relation varies, the slope remains constant across different epochs, suggesting that the different assembly mechanism acts similarly on both types of galaxies at different epochs. Moreover, the early type galaxies have a steep relation between mass-size, and they evolve faster with time. Whereas, late-type galaxies show a moderate evolution with time, as well as a shallow mass-size relationship, given as:
\begin{align}
\label{eqn:mass-size-early}
\begin{split}
Early-types:\\
R_e & \propto  M_*^{0.75} \ \ \ \ \ \ \ \ \ \ (\mathrm{for} \ M_*> 2\times 10^{10} \ \mathrm{M_\odot}),
\\
R_e & \propto (1+z)^{-1.48} \ \ \ (\mathrm{fast \ evolution}) 
\end{split}
\end{align}

\begin{align}
\label{eqn:mass-size-late}
\begin{split}
Late-types:\\
R_e & \propto  M_*^{0.22} \ \ \ \ \ \ \ \ \ \ (\mathrm{for} \ M_*> 3\times 10^{9} \ \mathrm{M_\odot}),
\\
R_e & \propto (1+z)^{-0.75} \ \ \ (\mathrm{moderate \ evolution}) 
\end{split}
\end{align}

We assess the quality of our dataset consisting of star-forming, disk-like galaxies (i.e., late-types) by comparing it with the above mass-size relation (Equation~\ref{eqn:mass-size-late}). As illustrated in the left panel of Figure~\ref{fig:scaling-Relations}, our dataset aligns well with the established relation. Utilizing the least-squares method of linear fitting, we obtain a slope of 0.22, which closely matches the value reported in \citet{Vanderwel2014}, and an intrinsic scatter of $0.13$ dex. This confirms the robustness of photometric quantities of our sample.

{\bf Evolution of Velocity Dispersion:}
The velocity dispersion of a galaxy is tightly coupled to its dynamical state and serves as an effective measure of turbulence. Its cosmic evolution can provide critical insights into the efficiency and nature of the underlying driving mechanisms, such as the baryonic feedback processes and gravitational interactions \citep[e.g.,][ and references therein]{ForsterSchreiber2006, Genzel2011, Swinbank2012, Newman2013, W15, Turner2017, HLJ17, Ubler2019}. Moreover, variations in the velocity dispersion with redshift could potentially suggest how galaxies interact with their environments, particularly the cosmic web \citep{Glazebrook2013}. The correlation between the ionized gas velocity dispersion and redshift is a well-established phenomenon, as reviewed comprehensively by \citet{Glazebrook2013} and \citet{Forster2020}.  In the right panel of Figure~\ref{fig:scaling-Relations}, we show this relation for our sample and compared with those of \citet{Ubler2019}. We infer that both datasets are in fair agreement across all redshifts with similar intrinsic scatter ($\approx 0.2$ dex). The slight offset in this relation can be attributed to the difference in the kinematic modeling techniques used in our analysis.




\section{Tully Fisher Relation}
\label{sec:TFR}
We assume that the galaxy masses (baryonic: stars and gas) scale with  the circular velocities as a power-law with slope ($\alpha$) and intercept ($\beta$), which can mathematically be defined as:
\begin{equation}
\label{eq:TFR}
\log(Y)=\beta + \alpha \ \log(X)
\end{equation}
where, $Y$ is the  list of stellar (or baryonic) masses and $X$ corresponds to the circular velocities ($V_c$). To obtain the best-fit to the data, we sample the likelihood using  Markov Chain Monte Carlo (MCMC), which uses an orthogonal likelihood,  defined as:
\begin{align} \label{eq:orth-ll}
        -2ln\mathcal{L} &=  \sum_{i} \ln(2\pi \sigma_i^2) +\sum_{i}\frac{(y_i-mx_i-b)^2}{\sigma_i^2(m^2+1)}\\
        {\rm where,} \
            \sigma_i^2 &= \frac{m^2\sigma_{x_i}^2+\sigma_{y_i}^2}{m^2+1} +\zeta_{int}^2
\end{align}
where, $x_i$ and $y_i$ denote the  stellar mass and circular velocity lists, respectively, while $\sigma_{x_i}$ and $\sigma_{y_i}$ represent their associated errors. The parameter $\zeta_{int}$ refers to the intrinsic scatter in the direction orthogonal to the best-fit line, and $\sigma_i$ gives the total scatter in the relation. We adopt this orthogonal likelihood fitting technique due to the significant scatter observed in \hz galaxies ($\sim 0.25$ dex) in both the stellar mass (at fixed velocity) and circular velocity (at fixed stellar mass). This scatter results in a dispersed relation, which is more accurately constrained by minimizing the scatter orthogonally along the best-fit line. This approach contrasts with the case of local disk galaxies, which exhibit a remarkably tight correlation in the $M_{star}-V_c$ (or $M_{bar}-V_C$) plane with a scatter of approximately $0.026-0.1$ dex on both the axes. In such cases, employing a vertical likelihood method (as described in Equation~\ref{eq:vert-ll1}) is more appropriate, as demonstrated by \citet{Lelli2019}. Further justification for the choice of orthogonal likelihood over vertical likelihood in the context of \hz data is provided in Appendix~\ref{app:TFR-FM}. Additionally, we remark that when fitting the STFR and BTFR, we utilize circular velocities calculated at $R_{out}$. As a result, the stellar and baryonic masses used in these fits are also constrained within the $R_{out}$ region, and denoted as $M_{star}$ and $M_{bar}$. However, wherever we use the total stellar or baryonic masses, they are denoted as $M_{star}^{Tot}$ or $M_{bar}^{Tot}$. For detailed discussion on the choice of global and constrained masses, we refer the reader to Appendix~\ref{app:TFR-total}.

In Figure~\ref{fig:TFR1}, we present the orthogonal likelihood fits for the STFR and BTFR, left and right panels respectively. For the STFR, we obtained a slope of $\alpha=3.03\pm 0.25$, an offset of $\beta = 3.34\pm 0.53$, and an intrinsic scatter of $\zeta_{int}=0.08$ dex. Correspondingly, the BTFR yielded $\alpha=3.21\pm 0.28$, $\beta=3.16\pm 0.61$, and $\zeta_{int}=0.09$ dex. These results are compared with previous studies, including both local and high-redshift studies of STFR and BTFR. For the STFR, we reference works by \citet[][$z\approx 0$]{Reyes2011}, \cite[][$z\approx 0$]{Lapi2018}, \citet[][$z\sim 1$]{ETD16}, \citet[][$0.9\leq z \leq 2.3$]{Ubler2017}, \citet[][$z\sim 1$]{AT2019a}, \citet[][$z\sim 0.9$]{Pelliccia2017}, \citet[][$z\sim 0.5-0.8$]{Abril21} and \citet[][$z\sim 2-2.5$]{Straatman17}. For the BTFR, we consider studies of \citet[][$z \approx 0$]{Papastergis2016}, \citet[][$z\approx 0$]{Lelli2019}, \citet[][$0.9\leq z \leq 2.3$]{Ubler2017}, \citet[][$z\approx 0$]{Goddy_2023}, \citet[][$z\sim 0.5-0.8$]{Abril21}, \citet[][$z\sim 0$]{Zaritsky2014-dm}, and \citet[][$z\sim 0$]{Catinella23}. As evident from Figure~\ref{fig:TFR1}, although previous studies of STFR and BTFR, both in locals and at high-redshift align well within $3\sigma$ uncertainties, the new data from \citet{GS23} offers evidence for a marginal evolution in both the slope and zero-point of these relations. Specifically, we report a slightly shallower slope and an increase in the STFR zero-point compared to most previous studies, as reported in Table~\ref{Tab:TFR-results}. 

\def\arraystretch{1.2}

\begin{table*}
\centering
\begin{tabular}{|c|c|c|c|c|c|c|c|}
\hline
Authors  & Redshift &  $\alpha$ &  $\beta$ & $\sigma_{int}$  & $\zeta_{int}$  &$\beta/\alpha$ \\
 &  & [$\log(km \ s^{-1})$] & [$\log(M_\odot)$]  & [dex] & [dex] & [zero-point]   \\
\hline
\multicolumn{7}{c}{\textbf{Stellar Tully Fisher Relation}}\\
\hline
\rowcolor{lightgray} 
This work & $z=0.6-2.3$ & $3.03 \pm 0.25$ & $3.34 \pm 0.53$ & 0.08 & 0.08 & 1.10\\
\hline
\citet{Reyes2011} & $z\approx 0$ & $3.80 \pm 0.01$ & $2.39 \pm 0.44$ & 0.11\textsuperscript{*} & 0.1 & 0.67\\
\hline
\citet{Lapi2018} & $z\approx 0$ & $3.67 \pm 0.23$ & $2.41 \pm 0.10$ & <0.1 &  0.12 & 0.66\\
\hline
\citet{ETD16} & $z\approx 1$ & $3.80 \pm 0.21$ & $1.88 \pm 0.46$ & -  & 0.09 & 0.49\\
\hline
\citet{AT2019a} & $z\approx 1$ & $3.70 \pm 0.30$ & $1.98 \pm 0.10$ & 0.17 &  0.16 & 0.54\\
\hline
\citet{Ubler2017} & $z=0.6-2.3$ & $3.60 \pm 0.01$ & $1.92 \pm 0.01$ & 0.22\textsuperscript{*} &  0.1 & 0.53\\
\hline
\setrow{\bfseries}\citet{Pelliccia2017} & $z\approx0.9$ & $3.68 \pm 0.79$ & $2.15 \pm 0.15$ & 0.11\textsuperscript{*} & 0.09 & 0.58\\
\hline
\setrow{\bfseries}\citet{Abril21} & $z=0.5-0.8$ & $4.03 \pm 0.63$ & $9.79 \pm 0.09$ & 0.43\textsuperscript{*} & 1.84 & 2.43\\
\hline
\setrow{\bfseries}\citet{Straatman17} & $z= 2-2.5$ & $5.18$ & $1.29$ & - & 0.66 & 0.25\\

\hline
\multicolumn{7}{c}{\textbf{Baryonic Tully Fisher Relation}}\\
\hline
\rowcolor{lightgray} 
This work  & $z=0.6-2.3$ & $3.21 \pm 0.28$ & $3.16 \pm 0.61$ & 0.09 & 0.09 & 0.98\\
\hline
\citet{Lelli2019} & $z\approx 0$ & $3.85 \pm 0.09$ & $1.99 \pm 0.18$ & 0.03-0.07 &  0.09 & 0.74\\
\hline
\citet{Papastergis2016} & $z\approx 0$ & $3.58 \pm 0.11$ & $2.33 \pm 0.01$ & 0.056 &  0.09 & 0.65\\
\hline
\citet{Goddy_2023} & $z\sim 0$ & $2.97 \pm 0.18$ & $4.04 \pm 0.41$ & -  & 0.13 & 1.36\\
\hline
\citet{Ubler2017} & $z=0.6-2.3$ & $3.73 \pm 0.10$ & $1.78 \pm 0.03$ & 0.23\textsuperscript{*} &  0.1 & 0.48\\
\hline
\setrow{\bfseries}\citet{Abril21} & $z=0.5-0.8$ & $3.50 \pm 0.20$ & $9.76  \pm 0.08$ & 0.25\textsuperscript{*} &  2.0 & 2.61\\
\hline
\setrow{\bfseries}\citet{Zaritsky2014-dm} & $z\approx0$ & $3.5 \pm 0.2$ & - & - & 1.77 & 1.26\\
\hline
\setrow{\bfseries}\citet{Catinella23} & $z\approx0$ & $3.06 \pm 0.08$ & $3.75\pm 0.17$ & 0.13 & 0.11 & 1.22\\
\hline

\end{tabular}
\caption{The slopes of  STFR and BTFR obtained in this work  along with a comparison to previous studies.
$\alpha$ and $\beta$ represent the slope and offset in the relation, respectively. $\sigma_{int}$ represents the intrinsic scatter from the respective studies and $\zeta_{int}$ denotes the orthogonal intrinsic scatter around the best-fit lines with respect to the GS23 dataset. The $\beta/\alpha$ value represents the zero-point of the relation.\\ \textsuperscript{*}These papers define scatter using the vertical distance between the data points and the best fit line.}
\label{Tab:TFR-results}
\end{table*}

\begin{figure}
		\includegraphics[width=\columnwidth]{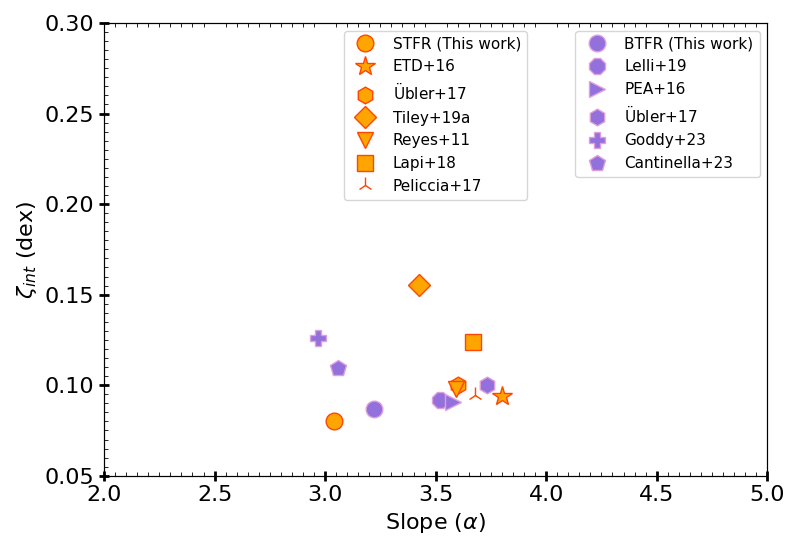}
		\caption{Comparison of slopes and intrinsic scatters: we take the best-fits of previous studies as a face-value (x-axis) and apply them on our dataset to compute the orthogonal intrinsic scatter (y-axis) around the adopted best-fit lines. STFR studies are represented by orange markers, BTFR studies by purple, with each marker corresponding to a distinct study listed in the legends. }
		\label{fig:scatter}
\end{figure}

\begin{figure*}
	\begin{center}
		\includegraphics[angle=0,height=8.5truecm,width=16.0truecm]{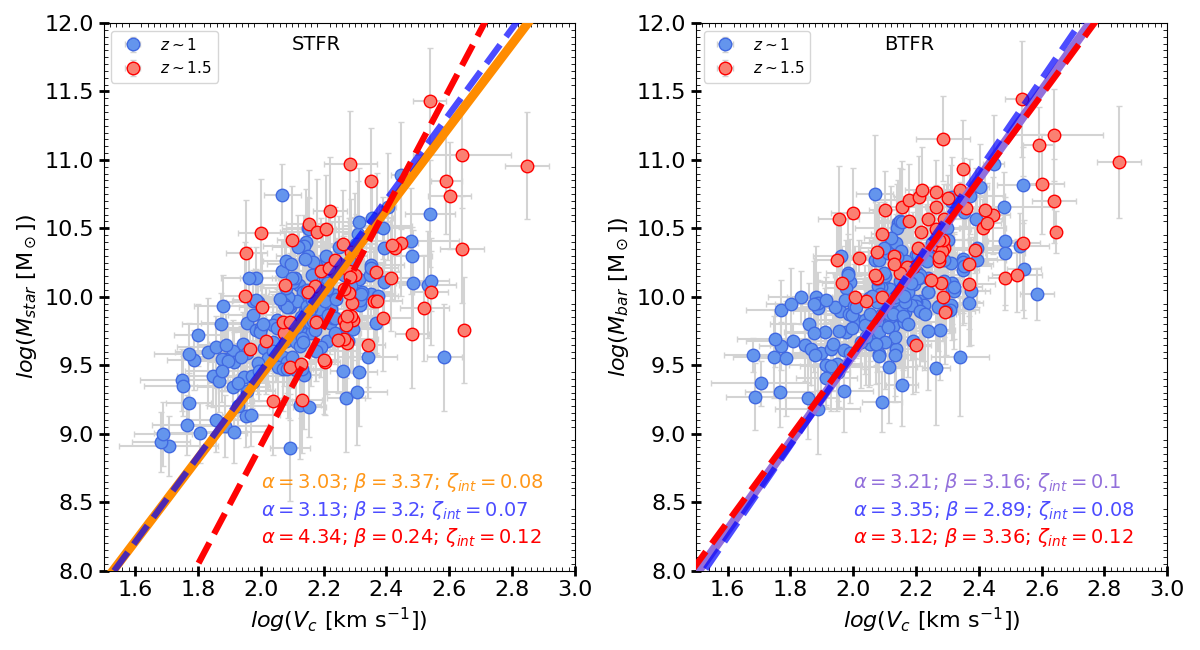}
		\caption{In the left and right panels, we present the STFR and BTFR, respectively, separated into two redshift bins: $0.6\leq z \leq 1.2$ ($\overline{z}\approx 1$) and $1.2 < z \leq 2.3$ ($\overline{z}\approx 1.5$). The bins corresponding to $z\sim 1$ and $z\sim 1.5$ are shown in blue and red respectively, and their respective fits are also displayed in red and blue colors. For reference, we have included the best-fits for STFR (in orange) and BTFR (in purple) derived from the full dataset. The associated best-fit parameters for each fit are provided at the bottom in their respective plots using the same color code as the best-fit lines.}
		\label{fig:TFR-z}
	\end{center}
\end{figure*}

\begin{figure*}
		\includegraphics[width=\columnwidth]{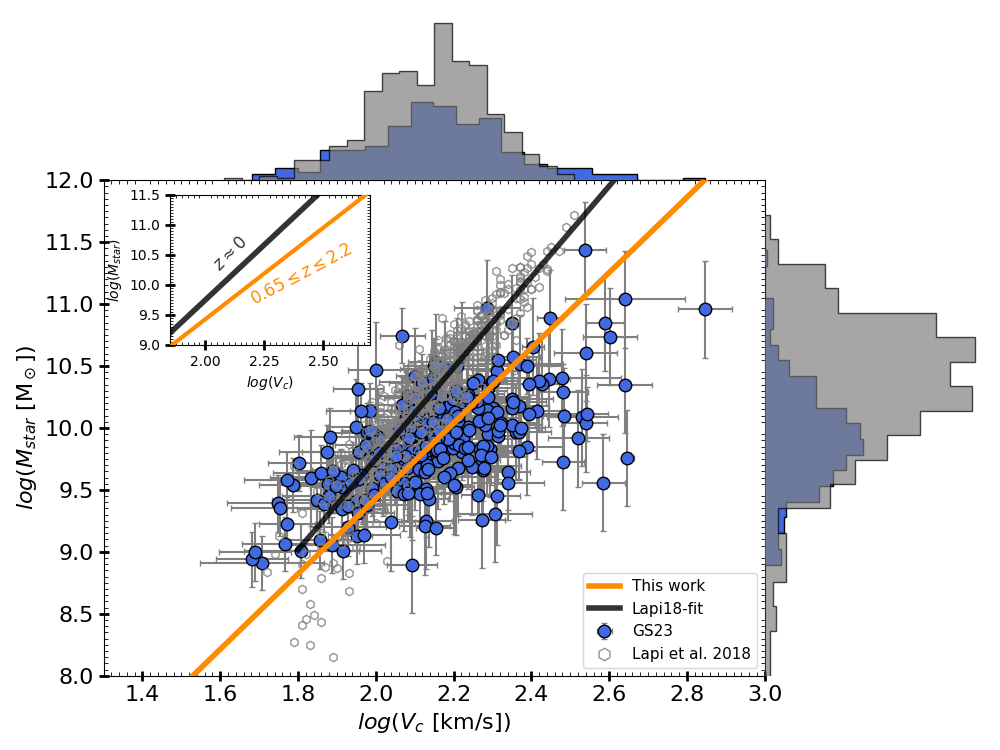}
  	\includegraphics[width=\columnwidth]{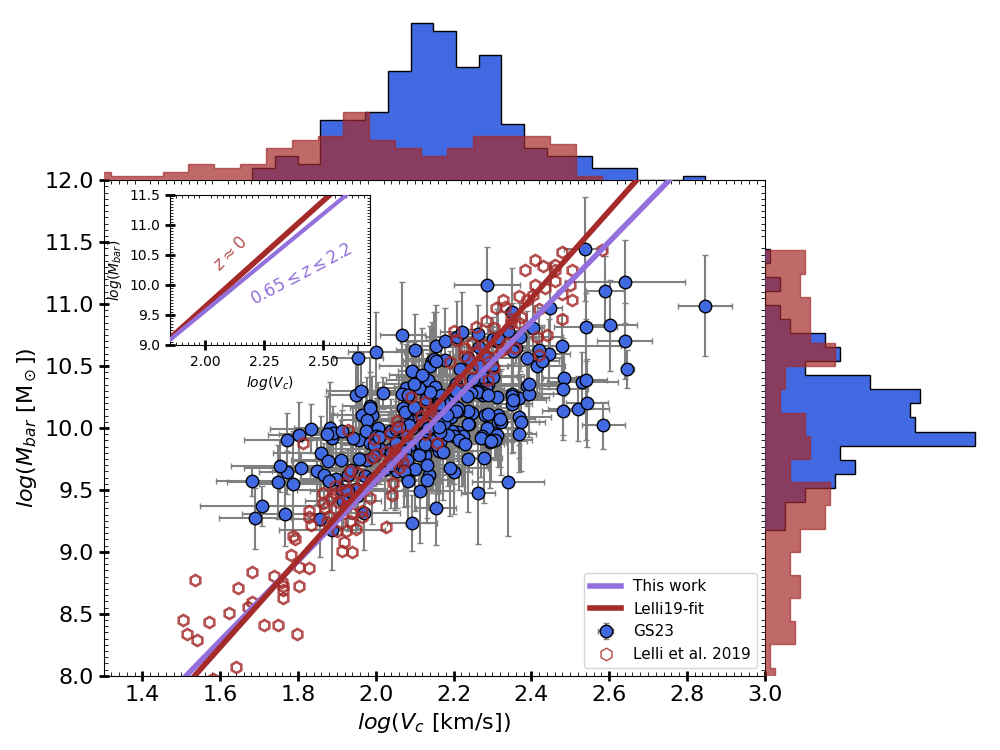}
		\caption{Comparison of STFR and BTFR with local studies. {\em Left Panel:} STFR comparison between the GS23 dataset (blue filled circles) and Lapi2018 (gray open circles). The best fit for Lapi2018 is represented by the black solid line, while the best-fit of this work (on GS23 data) is shown in orange. {\em Right Panel:} BTFR comparison between the GS23 dataset (blue filled circles) and Lelli2019 (brown open circles). The best fit for Lelli2019 is indicated by the brown solid line, while the best-fit of this work  in purple. In both panels, the inset provides a zoomed-in view of the local and high-$z$ fits within the range $9.0 \leq \log(M_{star/bar} \ [{\rm M_\odot}]) \leq 11.5$ and $1.65 \leq \log(V_c \ [{\rm km/s}]) \leq 2.75$. Additionally, for dataset comparison, histograms of the x and y axes are included: Lapi2018 in gray, Lelli2019 in brown, and GS23 in blue. In both the STFR and BTFR cases, we observe a divergent evolution in the slope, while the zero-point remains uncertain due to the absence of low-mass galaxies at high redshift.}
		\label{fig:TFR2}
\end{figure*}



Initially, we assumed that the observed change in the slope might be solely attributable to the fitting technique. To understand this, we fitted our data using the slope and zero-point values reported in the previous studies (listed in Table~\ref{Tab:TFR-results}) and calculated the intrinsic scatter around these reference lines. In Figure~\ref{fig:scatter}, we show the orthogonal intrinsic scatter as a function of the slopes obtained from prior studies, for both STFR (in orange) and BTFR (in purple). Our analyses indicates consistency with the slope and intrinsic scatter observed in previous studies. However, notably, our fitting technique yields shallower slope and reduced intrinsic scatter compared to previous studies, (see also  Table~\ref{Tab:TFR-results}). 

Although, previous studies have reported similar results, we place greater trust in our measurements. The reason is the outcome of a comparative analysis of orthogonal and vertical likelihood fitting techniques, as detailed in Appendix~\ref{app:TFR-FM}. In our study, we modeled the mock STFR data with high-$z$ errors on individual measurements and scatter, akin to observations at high redshifts. We observed that the vertical likelihood method could not retrieve the true slope at high-redshift, whereas the orthogonal likelihood method performed exceptionally well. We noted that the slope of the vertical likelihood differs by a factor of $1.5\pm 1$ compared to the orthogonal likelihood. Upon comparing our best-fit STFR/BTFR slopes with those of previous studies that minimize vertical scatter \citep[e.g.,][]{Reyes2011, Ubler2017}, we found the difference of factor $\sim 1.2$ to 1.5, similar to the one we just stated. Therefore, we suggest that orthogonal likelihood fitting techniques work best for high-$z$ datasets, which are prone to large scatter. Finally, we also fitted the STFR and BTFR using $V_{max}$, and observed that the slope only varies by about $\pm 0.15$ dex, which falls within the uncertainty range of the slope and zero-point provided using $V_c$. Based on these findings, we suggest that both the slope and the zero-point of the Tully-Fisher relation evolve modestly over cosmic time. Moreover, we learned that the slope, zero-point, and intrinsic scatter are all very sensitive to the preferred fitting technique.

 We would also like to note the reader that in this study, we do not focus on the evolution of the zero-point. This decision is based on our understanding that at high redshifts ($z$), the GS23 sample lacks low-mass galaxies due to Tolman surface brightness dimming \citep[for further details, refer to][]{GS23}, which are crucial for accurately constraining the zero-point of the TFR.

Furthermore, we explored the STFR and BTFR relations within different redshift bins, as shown in the left and right panels of Figure~\ref{fig:TFR-z}, respectively. In particular, we divided our galaxies into two redshift bins: $0.6\leq z \leq 1.2$ ($\overline{z}\approx 1$) and $1.2 < z \leq 2.3$ ($\overline{z}\approx 1.5$), fitting each bin independently using the aforementioned technique. Although Figure~\ref{fig:TFR-z} displays the best-fit results for both redshift bins, it is important to note that the $z\sim 1.5$ bin is biased towards massive galaxies and does not encompass the typical mass ($\log(M_{star/bar} \ [{\rm M_\odot}])=9.0-11.5$) and circular velocity ($\log(V_c \ {\rm km/s})= 1.6-2.85$) ranges upon which the fundamental TFR is established. Therefore, the results of the STFR and BTFR relations of $z\sim 1.5$ bin are not representative (or pertinent); hence we do not draw any conclusions for this redshift bin. Conversely, the STFR and BTFR relations at $z\sim 1$ cover typical mass and velocity range, and the fitting results are very similar to the one those derived from the full dataset. To be precise, for STFR, we find $\alpha=3.13, 
 \beta=3.20$,  and $\zeta_{int}=0.07$ dex, while for BTFR, we have $\alpha=3.35, \beta=2.89,$ and $\zeta_{int}=0.08$ dex. Thus, even when we restrict our analysis to galaxies at $z\sim 1$, we discern a nominal evolution in the slope and zero-point ($\beta/\alpha$) of the TFR relation at high-redshift.

\section{Discussion}
\label{sec:c6-discussion}
To reaffirm the validity of the GS23 dataset , which is a fair representative of the main sequence of star-forming galaxies as shown in Figure~\ref{fig:Mstar-sfr}, we further demonstrate its ability to accurately represent fundamental relations previously explored within similar redshift ranges using high-resolution photometry and resolved kinematics. Specifically, the mass-size relation \citep{Vanderwel2014} and cosmic evolution of the velocity dispersion \citep{Ubler2019} are shown in  the left and right panels of  Figure~\ref{fig:scaling-Relations}, respectively. It is evident from these figures that the dataset studied in \citet{GS23} is fairly representing these fundamental scaling relations, thereby reinforcing the robustness of  GS23 data and its suitability in studying the TFR. 

Moreover, the GS23 dataset spans stellar mass and circular velocity range as explored in local STFR studies. In particular, circular velocities ranges between $1.6 \lesssim \log(V_c \ [{\rm km/s}]) \lesssim 2.85$ and stellar masses $ 8.89 \lesssim \log(M_{star} \ [{\rm M_\odot}]) \lesssim 11.5$, which is the same range as explored in \citet{Reyes2011} and \citet{Lapi2018}. Therefore our sample is relatively free of selection bias (in terms of mass and velocity range) and hence allows us to study the STFR, as well as BTFR, as shown in Figure~\ref{fig:TFR1} left and right panels, respectively. We report a marginal evolution in the slope and zero-point of the STFR and BTFR relations for $z\leq 1$. Whereas, at $z\sim 1.5$ we do not draw conclusions on the evolution of the slope or zero-point due to insufficient data in the lower mass and velocity end. In subsequent sections, we discuss our results in light of previous local and \hz studies.

\subsection{Comparison with local studies}
To compare the STFR, we utilize the data from \citet[][hereafter Lapi18]{Lapi2018} as a benchmark. In the left panel of Figure~\ref{fig:TFR2}, we juxtapose the dataset of Lapi2018 with GS23. While the velocity ranges of both datasets overlap significantly, we observe that at higher velocities, local galaxies are more massive compared to their high-redshift counterparts. In other words, at fixed stellar masses (bench-marking against local galaxies), high-redshift galaxies exhibit fast rotation, a phenomenon also reported in previous studies such as \citet{Puech2008, Puech10, Cresci2009, Gnerucci2011, Swinbank2012b, Price16, AT2016, Straatman17, Ubler2017, Rizzo2020, Lelli2022, Lelli2023, GS23}. In particular, the slope and zero-point of the \hz STFR deviate from their standard values \citep[e.g.,][]{Lapi2018}  by approximately a factor of 1.2 and 0.72, respectively.
%
Specifically, in the redshift range $0.6\leq z \leq 2.5$, we obtain a slope of $\alpha=3.03\pm 0.25$ and an offset of $\beta = 3.34\pm 0.53$. Thus, we report a divergent evolution in the STFR over cosmic time. This marginal evolution is most-likely due the evolutionary stages of galaxies, which we plan to investigate in future work using cosmological simulations.

To compare the BTFR relation, we utilize the dataset from \citet[][hereafter Lelli2019]{Lelli2019} and juxtapose their data in the right panel of Figure~\ref{fig:TFR2}. Although GS23 dataset overlap seamlessly with Lelli2019, we notice that our dataset does not encompass galaxies with lower baryonic masses ($M_{bar}<10^{9.35} {\rm M_\odot}$) and lower velocities ($V_c < 40 {\rm km/s}$) as observed in local galaxies. This absence could be attributed to the Tolman dimming effect \citep{Tolman1930, Tolman_diming_1996}, as suggested in GS23. Due to these missing galaxies in the lower mass and velocity range, we refrain from making definitive conclusions regarding the zero-point of the BTFR at high-redshift. Secondly, similar to the STFR, at fixed baryonic mass, galaxies at \hz seems to rotate faster. Consequently, we observe a slightly shallower slope at \hz with respect to the locals.



\subsection{Comparison with high-$z$ studies} \label{sec:comp-high-z}
We acknowledge that \citet[][hereafter U17]{Ubler2017} and \citet[][hereafter AT19]{AT2019a} pioneered the study of the TFR at \hz using  large-datasets of IFU surveys: KMOS3D and KROSS surveys, respectively. Although, GS23 dataset consists of a sub-sample from both the KMOS3D and KROSS surveys, there exists discrepancies between the STFR and BTFR fits of U17, AT19, and this work. To understand these discrepancies, we present a comparative analysis of the STFR and BTFR datasets in Figure~\ref{fig:STFR-comp} and \ref{fig:BTFR-comp}, respectively. Moreover, we provide a detailed tailored comparison in Appendix~\ref{app:New-HZ-comp}. We remark the reader that AT19 only study the STFR, while U17 studies both STFR and BTFR. 

First, we note that the kinematic modeling techniques employed by the three studies are distinct. Differently to the approaches in A19 and U17 (see details in respective studies or briefly in Appendix C), GS23 fit the kinematics in 3D space. 
Some previous studies have shown that the 3D forward modeling allows for more accurate estimates of observed rotation velocities compared to 2D methods \citep{ETD16, GS21a, GS23}. In particular, the 2D kinematic modeling techniques overestimate the velocity dispersion and provide underestimated rotation velocities.
Consequently, the  discrepancies between these studies are expected. However, other factors that may contribute to these discrepancies are discussed below for each study separately:
\begin{figure}
		\includegraphics[width=\columnwidth]{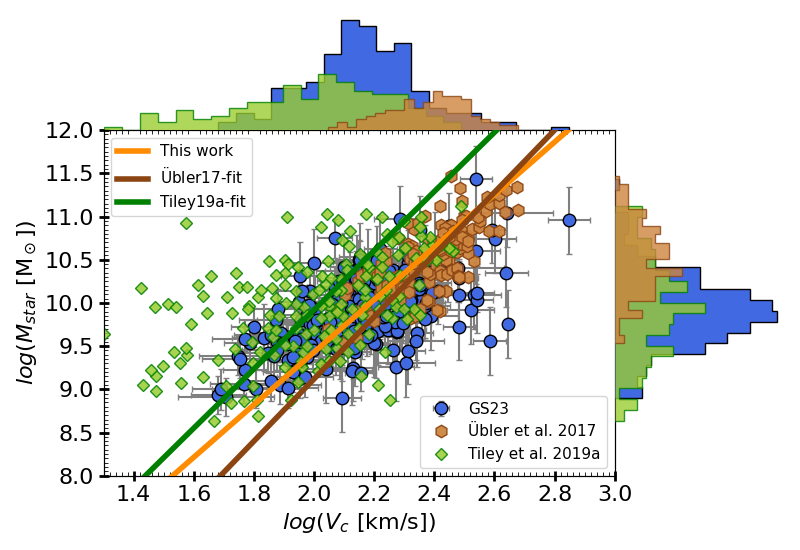}
		\caption{Comparison of STFR data of this work with datasets of other high-redshift studies: \citet[][green squares]{AT2019a} and \citet[][brown hexagons]{Ubler2017}. The orange, green, and brown lines represent the best-fit results for our study, \citet[][]{AT2019a}, and \citet[][]{Ubler2017}, respectively. To facilitate a comprehensive comparison of the datasets, we provide histograms showing the distributions of stellar mass and circular velocities, both horizontally and vertically, color-coded same as their respective datasets. We note that the velocities for ~\citet{AT2019a} are rotational velocities and not circular velocities.
		}
		\label{fig:STFR-comp}
\end{figure}

\begin{itemize}
    \item \textbf{AT19:} Rotation curves are derived along the major axis of the 2D velocity map, and beam-smearing corrections are applied only at the outer radius $(2-5 R_D)$. Moreover, the rotation curves were not corrected for the pressure gradients. Consequently, we anticipate lower circular velocity estimates in comparison to GS23. This is indeed evident in Figure~\ref{fig:STFR-comp}. The median value of the circular velocity distribution in the AT19 dataset is $\approx 100 \ {\rm km/s}$, whereas it is $\approx 150 \ {\rm km/s}$ in GS23, despite clear overlap of stellar mass distributions. Additionally, upon implementing the sample selection criteria ($V_c/\sigma > 3$) used in AT19 and utilizing rotation velocities (without pressure corrections), as illustrated in Figure~\ref{fig:Tiley-TFR-comp}, we still evident discrepancies in both the distributions and the best-fit results. Moreover, as shown in Figure~\ref{fig:scatter}, when we apply the AT19 best-fit to the GS23 dataset, we observe an intrinsic scatter of 0.16 dex, which is a factor of 2 higher than our estimates. Therefore, the AT19 fit is not applicable to the  GS23 dataset. Finally, we suggest that the observed differences between the best fits of AT19 and GS23 are primarily due to discrepant kinematic modeling methods and variations in fitting techniques. 
    \\

\begin{figure}
		\includegraphics[width=\columnwidth]{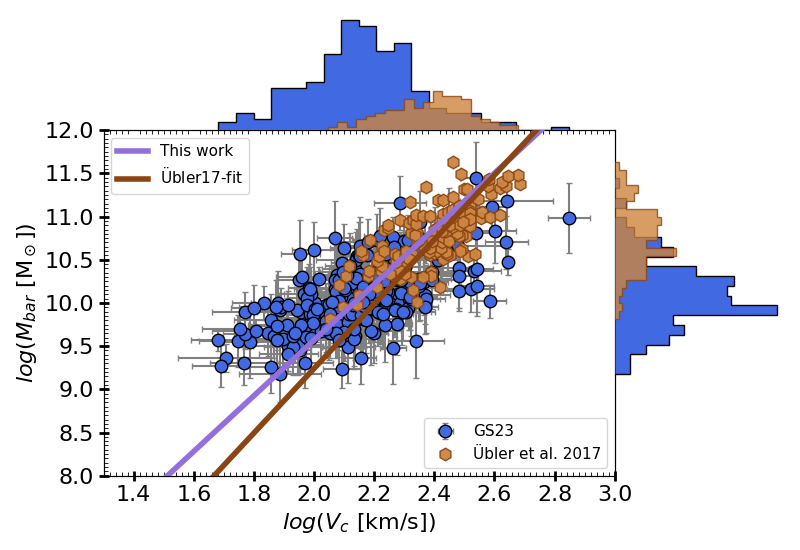}
		\caption{Comparison of BTFR dataset of this study with \citet[][brown hexagons]{Ubler2017}. The purple and brown lines represent the best-fit results for our study and \citet[][]{Ubler2017}, respectively. For a comprehensive comparison of the datasets, we provide histograms showing the distributions of stellar mass and circular velocities, both horizontally and vertically, color-coded correspond to their respective datasets.}
		\label{fig:BTFR-comp}
\end{figure}

    \item \textbf{U17:} Rotation curves are derived from the 2D velocity maps accounting for beam smearing and pressure gradient corrections. However, their pressure gradient corrections are applied under the assumption of constant and isotropic velocity dispersions. In \citet{GS21a} and \citet{Kretschmer2020}, it is shown that the  assumption of constant and isotropic velocity dispersion leads to overestimated circular velocities,  when corrected for pressure gradients, especially, in low rotation to dispersion ratio galaxies ($v/\sigma \lesssim 1.5$). Hence, circular velocity estimates of U17 are expected to be higher than GS23 estimates. It is indeed evident in Figures~\ref{fig:STFR-comp}, \ref{fig:BTFR-comp}, and \ref{fig:Ubler-TFR-comp}. 
   Even when we apply the sample selection cut ($V_c/\sigma > \sqrt{4.4}$) used in U17, we still observe high circular velocities at fixed stellar mass, as shown in Figure~\ref{fig:Ubler-TFR-comp}.
    In particular, we notice that U17 objects are biased towards higher velocities ($\overline{V_c}\approx 250 km/s$) and stellar/baryonic masses ($\overline{M_{star}} \approx 10^{10.5}M_\odot$ / $\overline{M_{bar}} \approx 10^{10.8}M_\odot$ ). While, GS23 dataset covers a typical mass and velocity ranges, we suggest that this is most-likely the primary reason for the  discrepant STFR and BTFR fits in U17 with respect to our work. Furthermore, when we apply the U17 best-fit to the GS23 dataset, we observe an intrinsic scatter of $\sim 0.1$ dex, which is a factor of 1.25 higher than our estimates. This suggest that discrepancies are also induced due to fitting techniques.
\end{itemize}

\subsection{Comparison of fitting techniques}
In this work, we performed a mock analysis of orthogonal and vertical likelihood fitting techniques on the stellar Tully-Fisher relation, as discussed in Section~\ref{sec:TFR} and detailed in Appendix~\ref{app:TFR-FM}. The mock analysis results are shown in Figures~\ref{fig:bestfit_sims} and \ref{fig:bestfit_simsz1}. We observe that the vertical likelihood fitting technique works well for the local Universe, where the intrinsic scatter is of the order of 0.01-0.1 dex. However, it underestimates the slope when intrinsic scatter exceeds 0.1 dex, as observed in the high-redshift data. In contrast, the orthogonal likelihood fitting technique performs best in both cases. Notably, it retrieves the correct slope with a precision error of less than $\pm 0.02$ dex. Consequently, we employed the orthogonal likelihood fitting technique in our work. The results of high-redshift STFR and BTFR fits obtained using orthogonal and vertical likelihood methods are presented in Figures~\ref{fig:TFR1} and \ref{fig:TFR1_vert}, respectively. Interestingly, the slopes obtained using the vertical likelihood for the STFR and BTFR differ from the orthogonal method's best fits by 1.72 dex and 1.98 dex, respectively. Therefore, we recommend using the orthogonal likelihood fitting technique for STFR and BTFR studies, or any scaling relations where data is subject to large uncertainties. For the reference of the reader, we have made our code publicly available via a \href{https://github.com/varenya27/Orthogonal-Fitting-Technique/tree/main}{GitHub repository}.

\section{Conclusions} 
\label{sec:c6-summary}
In this study, we investigated the Stellar and Baryonic Tully-Fisher relations over a broad redshift range of $0.6\leq z \leq 2.5$ using data from \citet{GS23}. To effectively address the substantial scatter prevalent among high-redshift galaxies, as elaborated in Appendix~\ref{app:TFR-FM}. We employed an orthogonal likelihood fitting technique, which minimizes the intrinsic scatter orthogonal to the best-fit line. The outcomes of our fitting methodology are presented in Figure~\ref{fig:TFR1}. For the STFR, our analysis yielded a slope of $\alpha=3.03\pm 0.25$, an intercept $\beta = 3.34\pm 0.53$, and an intrinsic scatter of $\zeta_{int}=0.08$ dex. Correspondingly, the best-fit BTFR parameters are: $\alpha=3.21\pm 0.28$, $\beta=3.16\pm 0.61$, and $\zeta_{int}=0.09$ dex. That is, the slopes of the STFR and BTFR are slower by a factor of $\sim 1.23$ and $\sim 1.15$, respectively, compared to those observed in the local Universe.

We also explored the relations for different redshift bins and found that the $z\sim 1.5$ bin was biased towards massive galaxies and hence inconclusive. Conversely, the $z\sim 1$ bin, devoid of such a bias, yielded results within the agreement to those derived from the complete (full) dataset, and affirmed the presence of minimal evolution in both the STFR and BTFR. When comparing our findings with local studies, we observed slight deviations, as shown in Figure~\ref{fig:TFR2}. Moreover, a comparison with previous \hz studies highlight differences due to kinematic modeling methods, fitting techniques, and sample selection, see Section~\ref{sec:comp-high-z} and Appendix~\ref{app:New-HZ-comp}. 

Through a comparative analysis of the outcomes obtained using orthogonal and vertical likelihood fitting methods, we have discerned a significant impact of fitting techniques on determining the slope and zero-point of scaling relations. Specifically, employing the vertical likelihood fitting technique at high redshifts (outlined in Appendix~\ref{app:TFR-FM}) led to a shallower slope of the STFR/BTFR by a factor of $\sim 2.5$, along with a correspondingly higher zero-point, as shown in Figures~\ref{fig:TFR1_vert}. This discrepancy arises from the inherent scatter within the observed data. Therefore, before picking a specific fitting technique, we suggest for conducting mock data analyses (including observed scatter) to evaluate the performance of different fitting techniques on given observations, as we demonstrated in Appendix~\ref{app:TFR-FM}.

Based on our findings, we conclude that the Tully-Fisher relation (TFR) exhibits a subtle shift in both the slope and zero-point values across cosmic time. This variation is most-likely due to dominant mechanisms driving galaxy evolution, such as gas accretion, star formation, mergers, or baryonic feedback. Therefore, we propose that the TFR is an empirical relation rather than a fundamental one in galaxy evolution, as it seems to show a dependency on galaxy's physical condition at a given epoch. Therefore, we emphasize the importance of studying the TFR across cosmic time using cosmological galaxy simulations to gain deeper insights into the underlying physical processes shaping the galaxy properties and its evolution across cosmic scales.

\begin{acknowledgements}
We thank the anonymous referee for constructive feedback. GS acknowledges the SARAO postdoctoral fellowship (UID No.: 97882) and the support provided by the University of Strasbourg Institute for Advanced Study (USIAS) within the French national programme Investment for the Future (Excellence Initiative) IdEx-Unistra. GS also thanks IIT Hyderabad for funding the January 2024 collaboration visit, which has led to this publication. 
\end{acknowledgements}

\bibliographystyle{aa} 
\bibliography{extracted.bib}

\begin{appendix}
\section{Fitting Techniques}\label{app:TFR-FM}

\begin{figure}[h]
    \centering
    \includegraphics[width=\columnwidth]{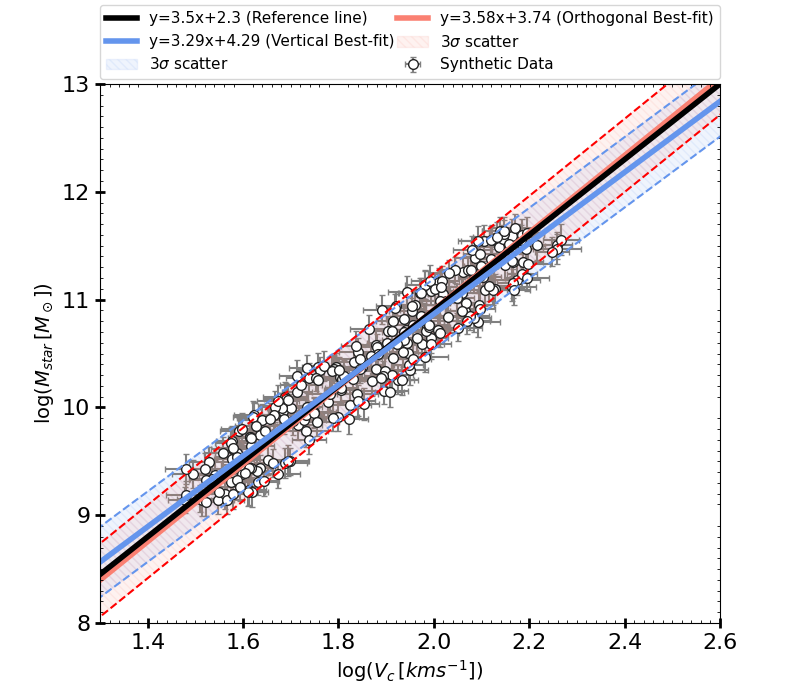}
    \includegraphics[width=\columnwidth]{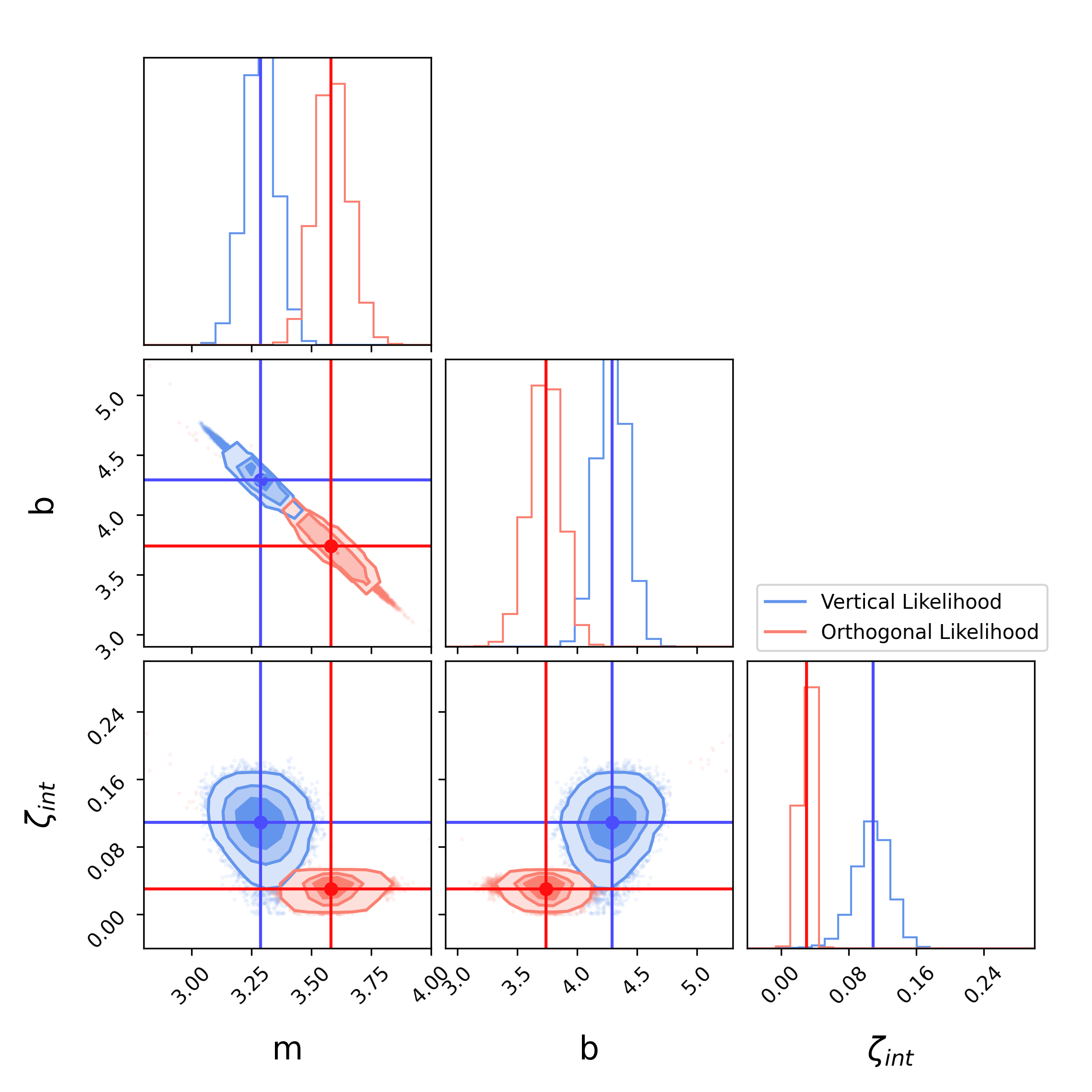}
    \caption{Vertical and orthogonal likelihood fitting on mock data resembling the local Stellar/Baryonic Tully-Fisher dataset. {\em Upper Panel:} The mock dataset is depicted as open black circles with errors, accompanied by the best-fit lines. The solid black line represents the true slope of $3.5$ used for generating the mock data. The blue and red solid lines correspond to the best fits obtained using the vertical and orthogonal likelihood methods, respectively. Both best-fit lines are followed by $3\sigma$ intrinsic scatter regions, indicated with the same color code as the best fits. {\em Lower Panel:} Posterior distributions resulting from MCMC-fitting for the vertical and orthogonal likelihoods are shown in blue and red, respectively. The contours within these corner plots illustrates the 68\%, 90\%, and 99\% credible intervals.
    }
    \label{fig:bestfit_sims}
\end{figure}

In this section we discuss the fitting techniques used in estimating the slopes and intercepts. We use the MCMC sampler \texttt{emcee} \citep{Foreman_Mackey_2013} to estimate the best-fit parameters. In our work, we fit a linear model $y=\alpha x+\beta$ to a set of $N$ datapoints $(x_i, y_i)$ with errors $(\sigma_{x_i}, \sigma_{y_i})$. We mainly consider two possible likelihoods for the parameter estimation. The first is a vertical likelihood that considers intrinsic scatter in the vertical direction, i.e., the intrinsic scatter varies only along the $y$-direction and not the $x$-direction \citep{Reyes2011, Lelli2019}. The second considers the intrinsic scatter to be orthogonal to the best-fit line \citep{Papastergis2016}. The vertical log-likelihood function follows:
\begin{align} \label{eq:vert-ll1}
            -2\ln\mathcal{L} &=  \sum_{i}^N \ln(2\pi \sigma_i^2) +\sum_{i}\frac{(y_i-\alpha x_i-\beta)^2}{\sigma_i^2}\\
                \sigma_i^2 &= {\alpha^2\sigma_{x_i}^2+\sigma_{y_i}^2} +\zeta_{int}^2\label{eq:scat_ver}
\end{align}
whereas, the orthogonal log-likelihood function is defined as:
\begin{align} \label{eq:orth-ll2}
        -2\ln\mathcal{L} &=  \sum_{i}^N \ln(2\pi \sigma_i^2) +\sum_{i}\frac{(y_i-\alpha x_i-\beta)^2}{\sigma_i^2(\alpha^2+1)}\\
            \sigma_i^2 &= \frac{\alpha^2\sigma_{x_i}^2+\sigma_{y_i}^2}{\alpha^2+1} +\zeta_{int}^2\label{eq:scat_ort}
\end{align}
where $\sigma_i$ in Equation~\eqref{eq:scat_ver} and \eqref{eq:scat_ort} encapsulate  the expressions for the total scatter in each case, calculated by taking into account the individual errors on the data points $(\sigma_{x_i}, \sigma_{y_i})$ along with the intrinsic scatter $\zeta_{int}$.

\begin{figure*}[h]
	\centering
	\includegraphics[width=\columnwidth]{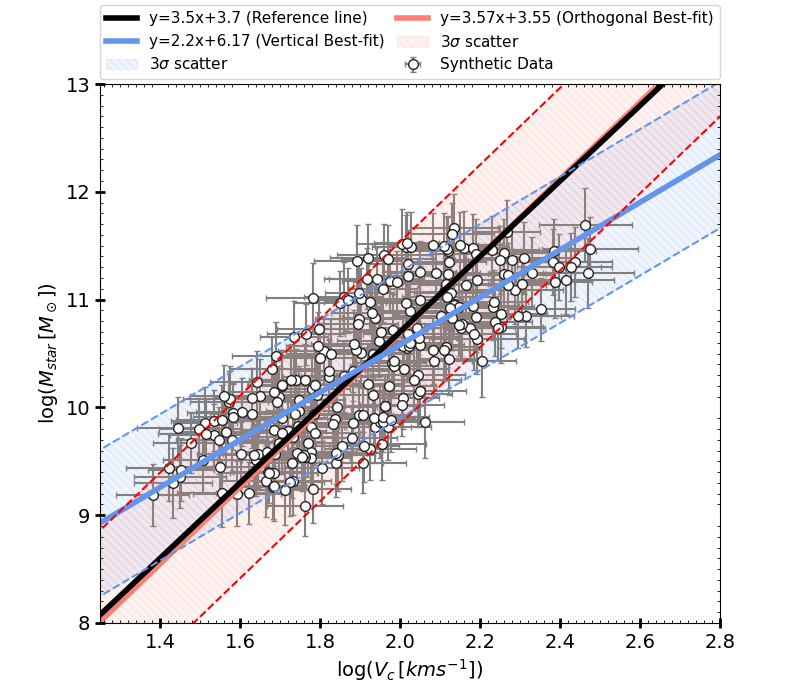}
	\includegraphics[width=\columnwidth]{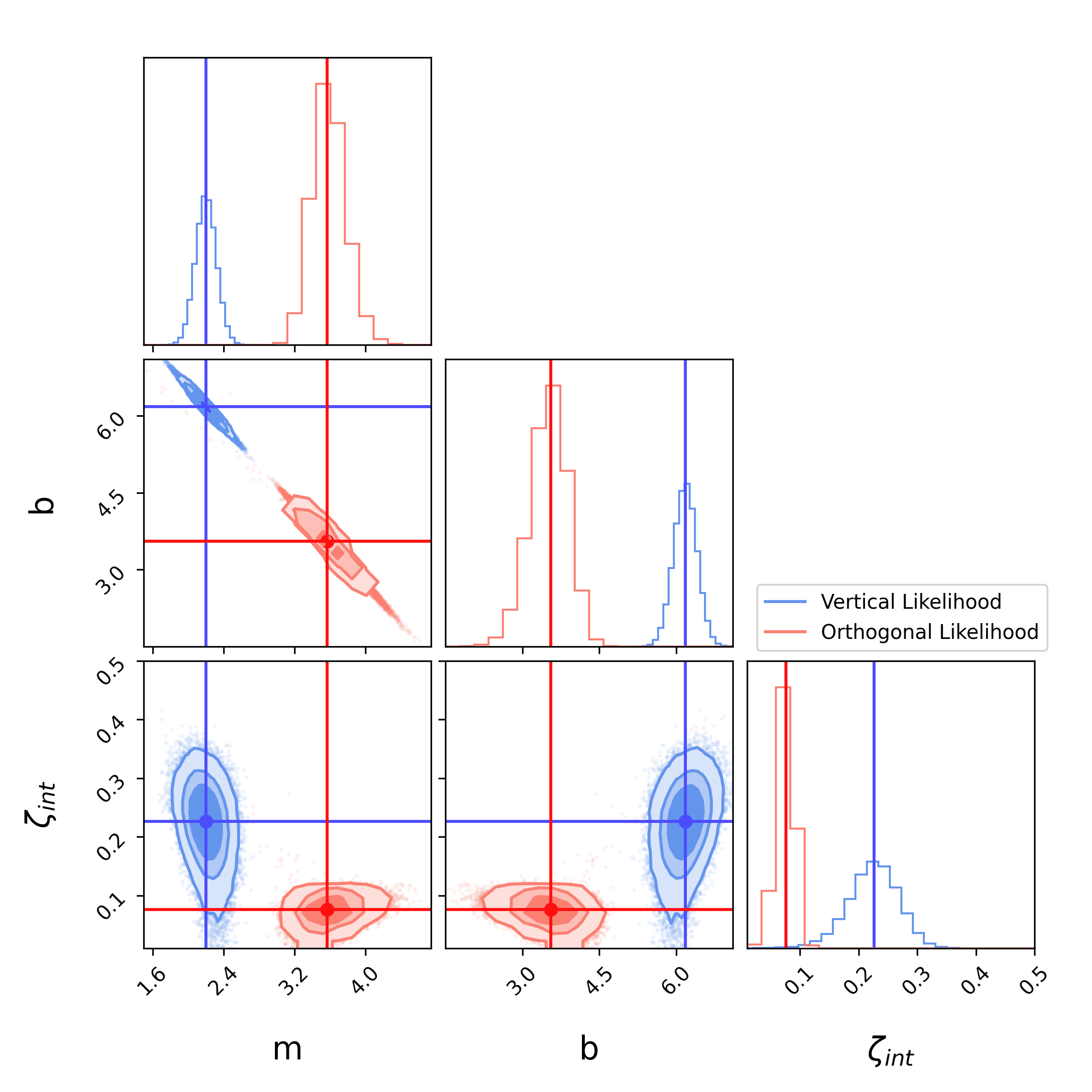}
	\caption{Vertical and orthogonal likelihood fitting on mock data resembling the \hz Stellar/Baryonic Tully-Fisher Relation. {\em Left Panel:} The mock dataset is depicted as open black circles with associated errors, accompanied by the best-fit lines. The solid black line represents the true slope of $3.5$ used for generating the mock data. The blue and red solid lines correspond to the best fits obtained using the vertical and orthogonal likelihood methods, respectively. Both best-fit lines are followed by $3\sigma$ intrinsic scatter regions, indicated with the same color code as the best fits. {\em Right Panel:} Posterior distributions resulting from the MCMC fitting process for the vertical and orthogonal likelihoods are shown in blue and red, respectively. The contours within these corner plots illustrates the 68\%, 90\%, and 99\% credible intervals.}
	\label{fig:bestfit_simsz1}
\end{figure*}

As it is not immediately evident which method is more suitable for our purposes, we use synthetic data to test the efficacy of both the methods by determining which of these likelihoods can recover the correct regression relation. First, we generate mock data using the commonly accepted value of the STFR, $\alpha=3.5$, and error values in the range of 0.12-0.15 and 0.03-0.08 for the $y_i$ (stellar mass) and $x_i$ (circular velocity) variables generated using a uniform distribution. The scatter (spread) in the  data-points is taken to be about 0.1 dex, which is an average value obtained  from previous studies \citep{Reyes2011, Lapi2018, Lelli2019}. Note that the scatter is applied in both, x and y, directions separately using a uniform distribution between -1 and 1. The mock dataset is shown in the top panel of Fig. \ref{fig:bestfit_sims}. The constraints ensure that median values of the errors and scatter are equivalent or higher than data in aforementioned literature. Subsequently, we fit this mock data using Bayesian inference with the {\tt emcee} sampler, employing both likelihoods, as illustrated in the bottom panel of Fig.~\ref{fig:bestfit_sims}. It is noteworthy that both likelihood functions reproduce the true values within  $1\sigma$ significance: $m_{VertL} = 3.22\pm0.06$ and $m_{OrthL}=3.54\pm0.08$. However, it is evident that the orthogonal likelihood closely constrains the true value.


Next, we assess the applicability of these likelihoods on mock data that closely mimics the high-redshift observations. Following a similar procedure as before, we generate mock data using the widely accepted value of the STFR, $\alpha=3.5$. However, in this case, we introduce a scatter of 0.25 dex, in both directions, consistent with the dataset of \citet{GS23, AT2019a, Ubler2017}, as shown in the left panel of Figure~\ref{fig:bestfit_simsz1}. The individual uncertainties on stellar masses and circular velocity measurements lie in the ranges 0.28-0.34 and 0.08-0.12 respectively, and they are applied using uniform distributions. Subsequently, we fit this data using both likelihoods, as shown in the right panel of Figure~\ref{fig:bestfit_simsz1}. Notably, the vertical likelihood recovers a slope of $2.20\pm0.12$, deviating by $1.30\pm 0.12$ from the true value (3.5), while the orthogonal likelihood accurately recovers the true value with nearly  100\% precision. Consequently, we propose that data exhibiting higher scatter, as often observed at high redshifts, necessitates advanced fitting techniques, such as an orthogonal likelihood which minimizes the intrinsic scatter perpendicular to the best-fit. Thus, in this work we employ the orthogonal likelihood to estimate the best-fit for the STFR and BTFR at high-redshift.



\section{STFR and BTFR with Total Masses}\label{app:TFR-total}
\begin{figure*}[h]
    \centering
    \includegraphics[width=\columnwidth]{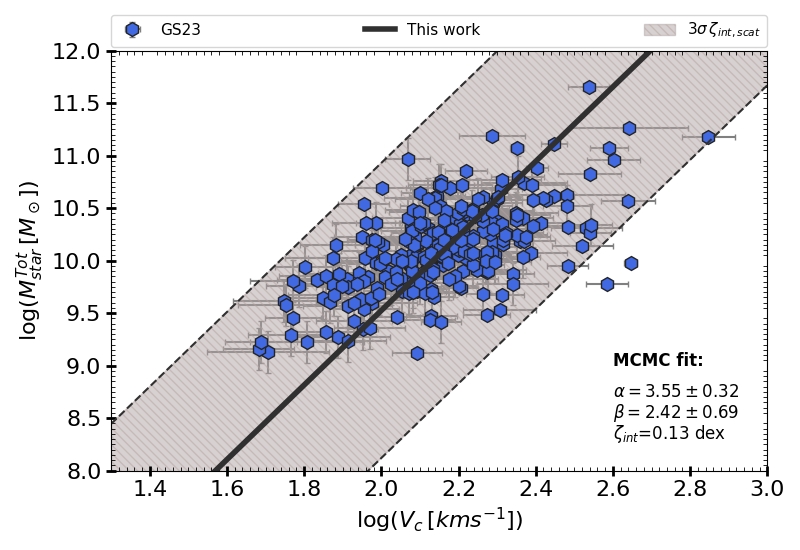}
    \includegraphics[width=\columnwidth]{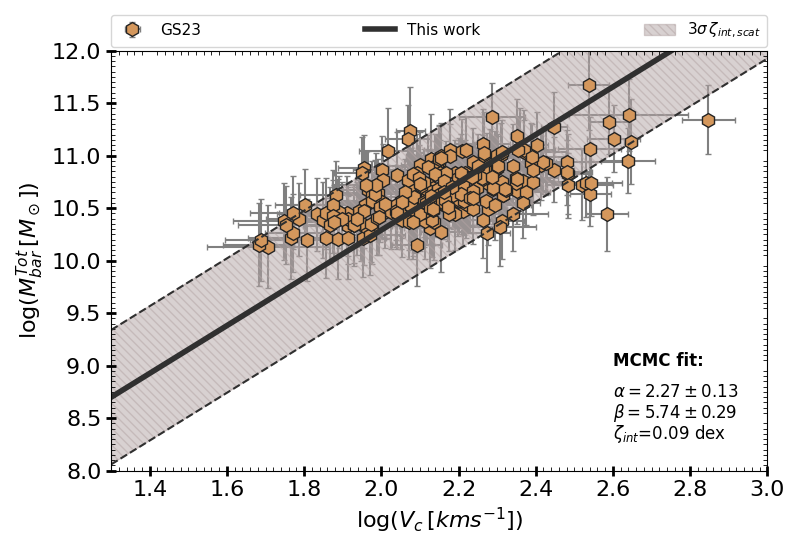}
    \caption{STFR and BTFR computed using the total masses instead of the stellar and baryonic masses contained within $R_{out}$. The color codes are given in the legend of the plot.}
    \label{fig:TFR-full-mass}
\end{figure*}

In Section~\ref{sec:TFR}, while computing the Tully-Fisher relations, we use the stellar and baryonic masses of the galaxies contained within $R_{out}$. In the rotation curves for galaxies at low redshifts ($z\sim0$), we observe a clear maximum value of the velocity, followed by a flat curve. In such cases, it is therefore more meaningful to use the velocity of the flat portion along with the total mass (stellar or baryonic) or luminosity in the study of Tully-Fisher Relation. At high redshifts, however, since we do not observe a substantial portion of the curve flattening in all galaxies, we cannot be sure if the $V_{max}$ measurements indeed represent the maximum circular velocity of the galaxies, as apparent in Figure~\ref{fig:Vc-Vmax}. Therefore, to maintain uniformity across the sample, to treat all galaxies consistently, and facilitate comparisons with previous \hz\ studies \citep[e.g.,][]{AT2019a, Ubler2017}, we use the circular velocity computed at $R_{out}$. Consequently, in Section~\ref{sec:TFR} for STFR and BTFR, we use stellar and  baryonic masses computed within $R_{out}$. However, in Figure~\ref{fig:TFR-full-mass}, we present the STFR and BTFR using the total stellar and baryonic mass ($M_{star}^{Tot}$ and $M_{bar}^{Tot}$) of a galaxy, employing the same techniques as described in Section~\ref{app:TFR-FM}. For the STFR, we find a slope of $\alpha = 3.55\pm0.32$, an offset of $\beta=2.42\pm0.69$, and an intrinsic scatter of $\zeta_{int}=0.13$ dex. In the case of the BTFR, we observe a slope of $\alpha = 2.27\pm0.13$, an offset of $\beta=5.74\pm0.29$, and an intrinsic scatter of $\zeta_{int}=0.09$ dex. It is noteworthy that the STFR maintains nearly the same slope (steeper by 0.52 dex) and offset, as when using the stellar mass within $R_{out}$ (see Figure~\ref{fig:TFR1}). However, the slope and offset of the BTFR notably differ, shallower by 0.94 dex and higher by 2.58 dex respectively.

The results of BTFR are rather surprising as they suggest that low-mass galaxies at high redshifts are highly gas-dominated systems. The later is previously suggested in \citet{Tacconi2020}. However, given the shallowness of the relation, we underscore the need for accurate estimates of gas mass at high redshift. Most likely, the limitations lie in the HI scaling relations, which appear insufficient to constrain the total HI mass at high redshift. For these reasons, we do not report STFR and BTFR from the total mass in the main text. However, it could also be attributed to the lack of deep observations, potentially resulting in incomplete mapping of the circular velocity in low-mass galaxies, as reported in the latest study of \citet{GS23}. Most likely, high-resolution (deep integration time) observations are required to put tighter constrains on TFR at high redshift. 


\section{Tailored comparison of STFR and BTFR with high-$z$ studies} \label{app:New-HZ-comp} 

As discussed in Section~\ref{sec:c6-data}, the GS23 sample represents a rotation-supported system, lies on and around the main sequence of star-forming galaxies, and adheres to fundamental scaling relations (see Figure~\ref{fig:scaling-Relations}). This suggests that the TFR of the full GS23 sample can be directly compared with local TFR studies as shown in Figure~\ref{fig:TFR2}. As well as to keep the consistency towards sample selection between the local and the high-z Universe, we compare TFR of full GS23 sample with previous high-$z$ studies as shown in Figure~\ref{fig:STFR-comp} and \ref{fig:BTFR-comp}.

However, it is evident from the works of previous high-$z$ studies, namely \citet{AT2019a} and \citet{Ubler2017}, that their sample selection, measurements of stellar and gas mass, and fitting techniques differ from those employed in local studies, as well as being discrepant from our approach. Therefore, in this section, we briefly discuss the physical properties of the aforementioned studies that are relevant to the TFR, emphasize their sample selection, and then compare their TFR fits with our datasets. In particular, \citet{AT2019a} study is compared with KROSS sample and \citet{Ubler2017} with KMOS3D sample.

\begin{figure*}[h]
    \centering
    \includegraphics[width=\columnwidth]{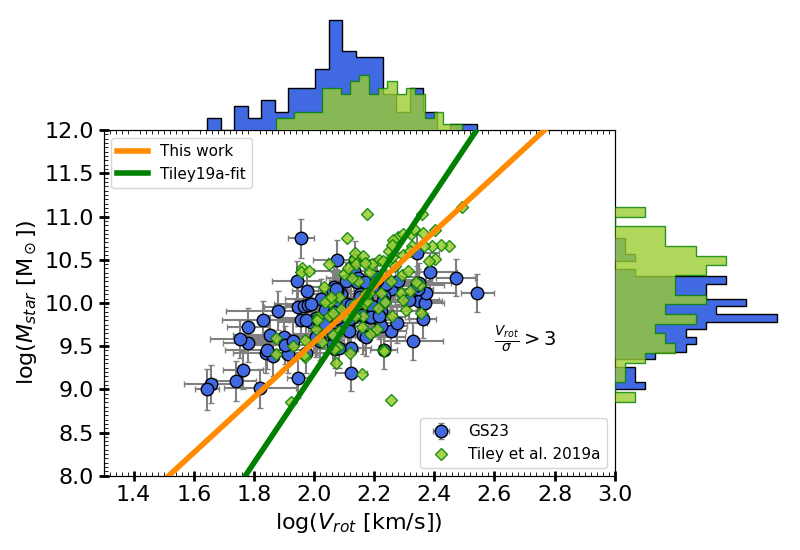}
    \includegraphics[width=\columnwidth]{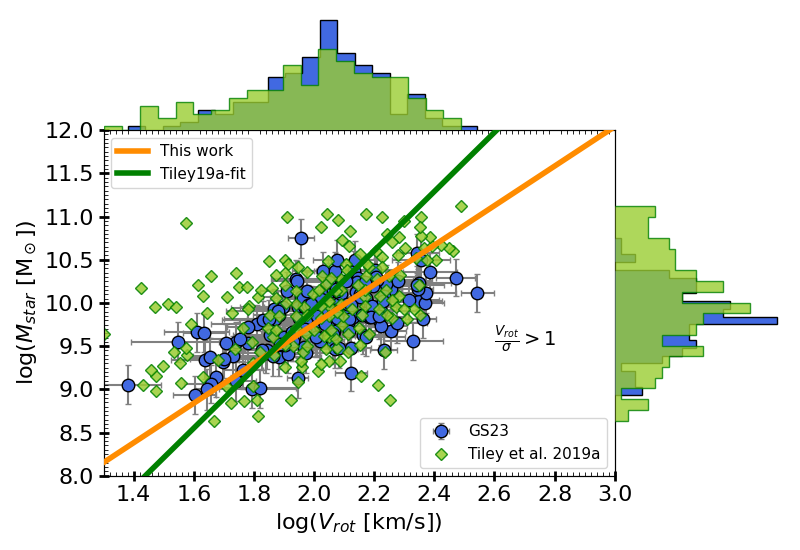}   
    \includegraphics[width=\columnwidth]{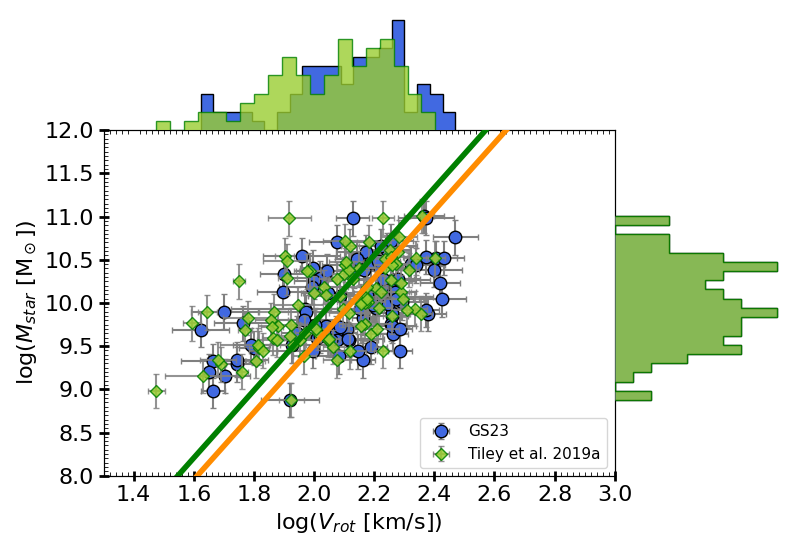}   
    \caption{Comparison of the STFR, using only KROSS sample studied in GS23 datasets, with \citet{AT2019a}. The orange and green lines represent the best fit lines for our and \citet{AT2019a}'s work, respectively. The blue and green data points represent- 	\textit{Upper Left Panel:} disk-like galaxies characterized by $V_{\rm rot}/\sigma>3$,  \textit{Upper Right Panel:} rotation supported galaxies characterized by $V_{\rm rot}/\sigma>1$, and \textit{Lower Panel:} KROSS and \citet{AT2019a} matching sample for $V_{\rm rot}/\sigma>1$. We note the reader that in this comparison, we use rotation velocity with in $R_{\rm out}$ (i.e., $V_{\rm rot}$), which is not corrected for pressure support, i.e., $V_{\rm rot}$ is comparable to $V_{2.2}$ of \citet{AT2019a}. }
    \label{fig:Tiley-TFR-comp}
\end{figure*}


\begin{figure*}[h]
    \centering
    \includegraphics[width=\columnwidth]{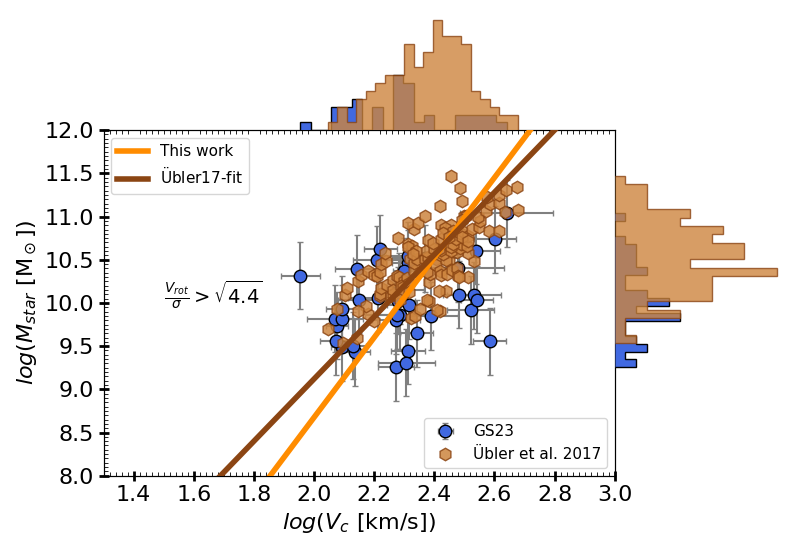}
    \includegraphics[width=\columnwidth]{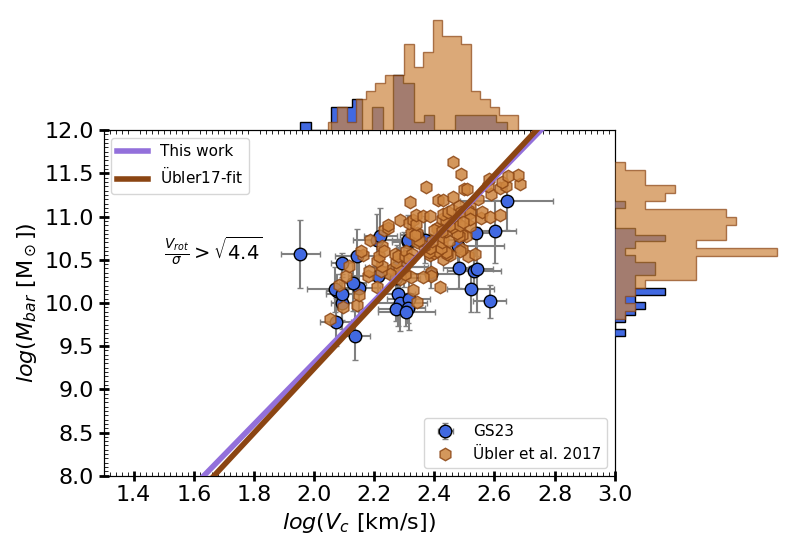}
    \includegraphics[angle=0,height=6.0truecm,width=9.0truecm]{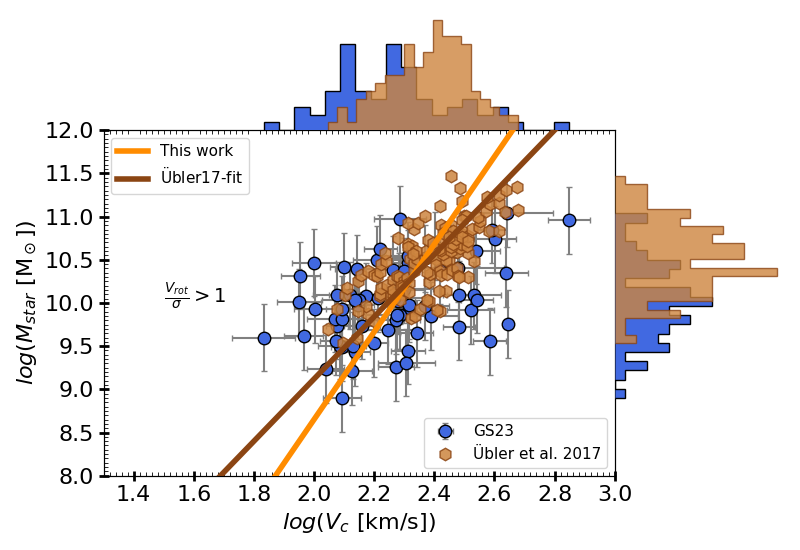}
  	\includegraphics[angle=0,height=6.0truecm,width=9.0truecm]{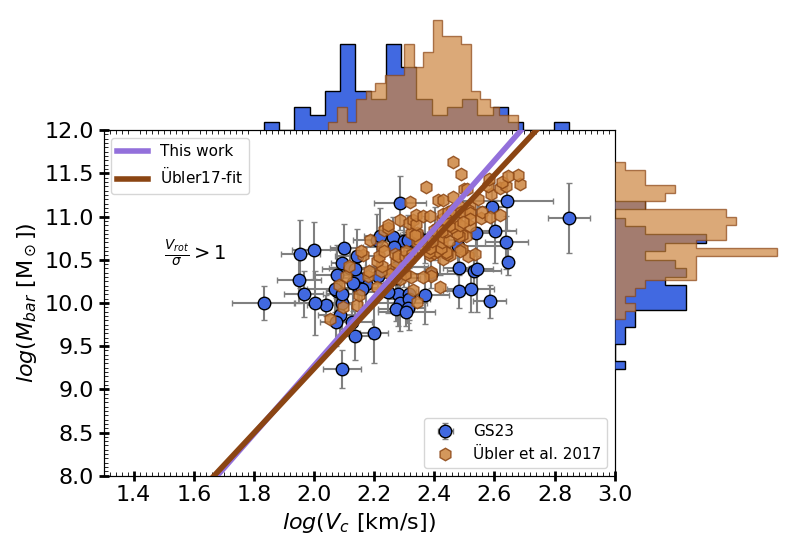} 
        \caption{Comparison of the STFR and BTFR, using only KMOS3D sample studied in GS23 datasets, with \citet{Ubler2017}. In the top panel figures, we restrict the analysis for $V_{\rm rot}/\sigma>\sqrt{4.4}$, whereas in the bottom panels, we analyze the full KMOS3D sample studied in GS23 for $V_{\rm rot}/\sigma > 1$. \textit{Left Panels:} STFR comparison between the two studies. The orange and brown lines represent the best fit lines for our and \citet{Ubler2017} work respectively. \textit{Right Panels:} BTFR comparison between the two studies. The purple and brown lines represent the best fit lines for our and \citet{Ubler2017} work respectively.}
    \label{fig:Ubler-TFR-comp}
\end{figure*}

\textbf{\underline{Comparison with Tiley+19:} }
Following \citet{H17}, Tiley et al. constructed line-of-sight velocity maps using $H_\alpha$ emission. To determine the rotation velocity of the system, a $0.7"$ slit was placed along the major axis of the velocity map, enabling them to determine the rotation velocity using dynamical modeling based on the arctangent disk model \citep{Courteau1997}. Subsequently, \RCs\ were corrected for beam-smearing using the prescription outlined  in \citet{HLJ17}. We note that their \RCs\ are not corrected for pressure support; therefore, when comparing STFR, we use the rotation velocity of the GS23 sample instead of circular velocity (which incorporates pressure support corrections). The stellar masses of AT19 sample are derived using SED-fitting with the \texttt{LEPHARE} code \citep{Arnouts1999, Ilbert2006} as discussed in \citet{AT2019a}. In this work, we utilize the same stellar masses received through private correspondence with Alfred Tiley. The study by \citet{AT2019a} focuses only on STFR for two distinct cases: (1) a disky sample characterized by $V_{\rm rot}/\sigma > 3$, and (2) a rotation-supported sample, i.e., $V_{\rm rot}/\sigma > 1$. We compare our sample and its best fit to both cases, as illustrated in Figure~\ref{fig:Tiley-TFR-comp}, with the upper left and right panels representing the disky and rotation-supported samples, respectively. 

In case of disky sample, upper left panel of Figure~\ref{fig:Tiley-TFR-comp} , we notice that AT19 sample is dominant towards massive systems (hence fast rotating). Conversely, the GS23 sample contains a larger dynamic range in velocities, i.e., also the lower velocities (low masses). This discrepancy is most-likely due to the capability of 3D forward modeling to account for low-mass systems, which are often overlooked in 2D kinematic modeling. Our analysis yields a slope of $3.20\pm0.48$, diverging by a factor of 2.0 from the slope observed in AT19. This difference is yet again attributed to variances in kinematic modeling and fitting methodologies, which likely contribute to the shallow slope and distinct offset in our work.

In the case of rotation-supported systems (upper right panel), the distributions of both samples match very closely in terms of velocity and stellar masses, as expected. However, there is a notable discrepancy in the best-fit  relation. In particular, our slope of $2.29\pm0.22$ differs from \citet{AT2019a} by a factor of 1.14 dex. To understand this discrepancy, we matched the AT19 and KROSS samples from \citet{GS21a}, finding 96 matches, as shown in the bottom panel of Figure~\ref{fig:Tiley-TFR-comp}. We noticed that the discrepancy is only in terms of velocity estimates. When we fit this matched data of AT19 and GS23 using the orthogonal likelihood fitting technique, we obtained similar slopes that differ only in terms of offset due to differences in velocity estimates. This suggests that the discrepancy in the slope between AT19 and this work arises mainly from the fitting techniques. However, a minor difference could also stem from variations in beam-smearing corrections, which are implemented differently in both studies, or from the circular velocity measurements which are taken at different radii.


\textbf{\underline{Comparison with Ubler+17.:}} In accordance with \citet{Wuyts2016}, \citet{Ubler2017} determined the radial velocity and velocity dispersion of the KMOS3D sample by placing a circular aperture with a diameter of $0.8\arcsec$ along the kinematic major axis, utilizing the LINEFIT code \citep{Davies2009}, which considers spectral resolution. To obtain circular velocity profiles (rotation curves) for the system, they employed dynamic mass modeling of kinematic profiles using DYSMAL \citep{Cresci2009, Davis2011}, allowing for an exponential disk with a Sersic index of $n_s =1$. This modeling procedure encompasses the coupled treatment of radial velocity and velocity dispersion, incorporating beam-smearing \citep[see,][]{Davies2009, Davis2011} and pressure support correction assuming constant and isotropic velocity dispersion \citep{Burkert2010}.

It is crucial to note that \cite{GS23} employs 3DBarolo to estimate circular velocity profiles, utilizing a non-parametric approach while simultaneously correcting for spectral and spatial resolution (i.e., beam-smearing) in 3D space. Subsequently, these rotation curves (\RCs) underwent pressure support corrections, as detailed in \citet{GS21a}, following the methodology of \citet{Anne2008}. In the latter, pressure support corrections do not assume constant and isotropic velocity dispersion unlike \citet{Burkert2010}; instead, they account for velocity anisotropies. For further details we refer the reader to \citet{GS21a} and \citet{GS23}. 

The star-formation rates and stellar masses in \citet{Ubler2017} are estimated through proper SED fitting techniques discussed in \citet{Wuyts2011b, W15}, and \citet{W19}. Molecular mass estimates are obtained using the scaling relation of \citet{Tacconi2018}. The HI gas mass is considered negligible within $1-3 R_e$, i.e., $M_{\rm bar} = M_{\star} + M_{\rm H2}$. However, following \citet{Burkert2016}, the author applied larger uncertainties ($0.2$ dex) to total gas mass measurements to account for missing HI mass. 
In comparison, \citet{GS23} consider $ M_{\rm bar} = M_{\star} + M_{\rm H2} + M_{\rm HI}$, where the estimates for stellar and molecular gas mass align with those of \citet{Ubler2017}. However, the HI mass is derived using a scaling relation. Further details on baryonic mass estimates are provided in Section~\ref{sec:Mbar}.

In terms of sample selection, \citet{Ubler2017} focuses on galaxies on-and-around the main-sequence of star-forming galaxies. However, to select the most disk-like systems, they apply a $V_{\rm rot}/\sigma > \sqrt{4.4}$ cut and then study the STFR and BTFR. To facilitate one-to-one comparison of our fitting techniques, we select only KMOS3D sample from GS23 and apply a $V_{\rm rot}/\sigma > \sqrt{4.4}$ cut, resulting in 53 remaining galaxies. We show the results of this tailored comparison in Figure~\ref{fig:Ubler-TFR-comp}, with the STFR in the left panel and the BTFR in the right panel. Notably, the distribution of stellar/baryonic mass and circular velocities is skewed in our sub-sample. In contrast, the \citet{Ubler2017} sample comprises 135 galaxies with a Gaussian distribution in both mass and velocities.
In the GS23 sub-sample, the median circular velocity is approximately 150 km/s, with a stellar mass of $\sim 10^{9.7} M_\odot$ and a baryonic mass of $\sim 10^{10.3} M_\odot$. Conversely, in the \citet{Ubler2017} sample, the median circular velocities are around 250 km/s, with stellar and gas masses at $\sim 10^{10.5} M_\odot$. For this tailored comparison, our sample is biased toward intermediate-mass systems, while the \citet{Ubler2017} sample consists mainly massive systems.

To fit the STFR and BTFR of the GS23 KMOS3D sub-sample, we employ the same techniques established in Appendix~\ref{app:TFR-FM} and applied in Section~\ref{sec:TFR}. The results are presented in Figure~\ref{fig:Ubler-TFR-comp}. For the STFR, we report a slope of $\alpha = 5.07^{+1.09}_{-1.17}$, intercept $\beta = -1.49^{+2.64}_{-2.28}$, intrinsic scatter of 0.15 dex. For the BTFR, the reported values for the  slope is $\alpha = 3.96^{+0.81}_{-1.24}$, intercept corresponds to  $\beta = 1.36^{+2.81}_{-1.84}$,  and an intrinsic scatter of 0.14 dex. 

It is evident that at fixed stellar or baryonic mass, the  circular velocity in \citet{Ubler2017} sample is higher by factors of 1.3-1.5. This is most-likely due to their pressure support correction method that assumes constant and isotropic velocity dispersion, which overestimates the circular velocity across the galactic scales. Additionally, we compare the STFR and BTFR of full KMOS3D sample of GS23 with \citet{Ubler2017} in second row of Figure~\ref{fig:Ubler-TFR-comp}. Nevertheless, we encountered the similar evolution in TFR slopes. This suggests that differences are arising due to difference in kinematic modeling and fitting techniques employed in our work.

\section{Fits with vertical scatter}
\begin{figure*}
		\includegraphics[width=\columnwidth]{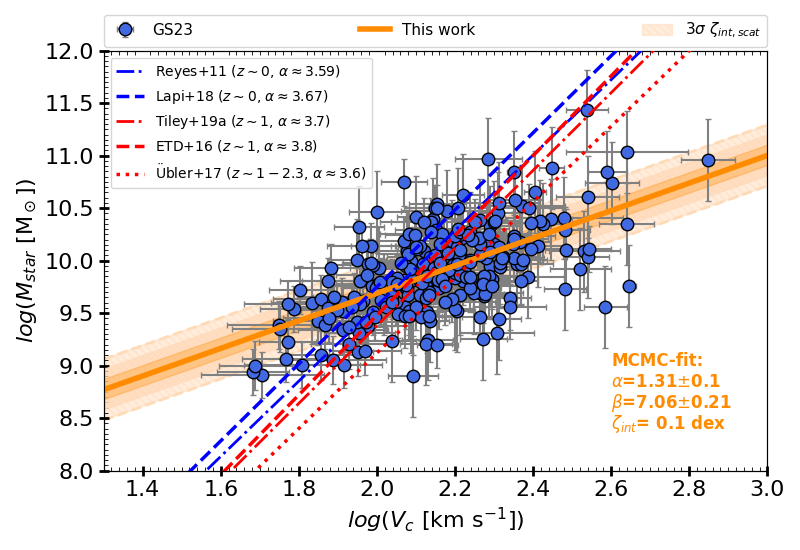}
		\includegraphics[width=\columnwidth]{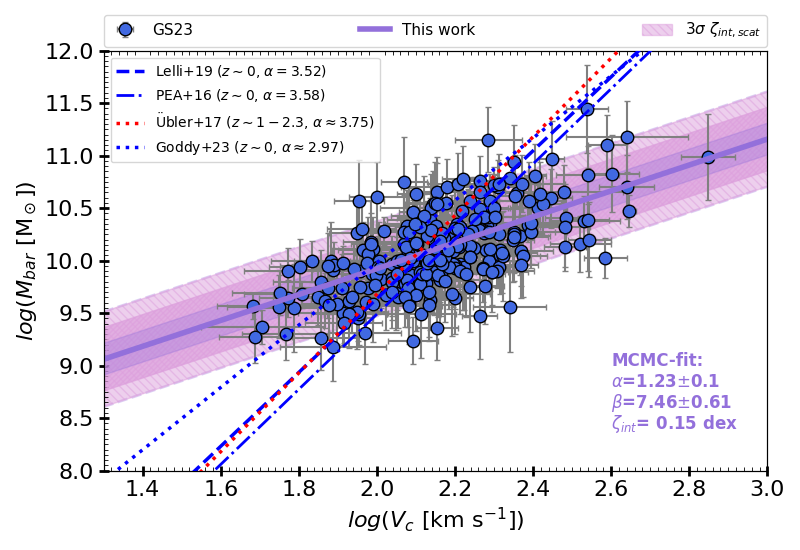}
		\includegraphics[width=\columnwidth]{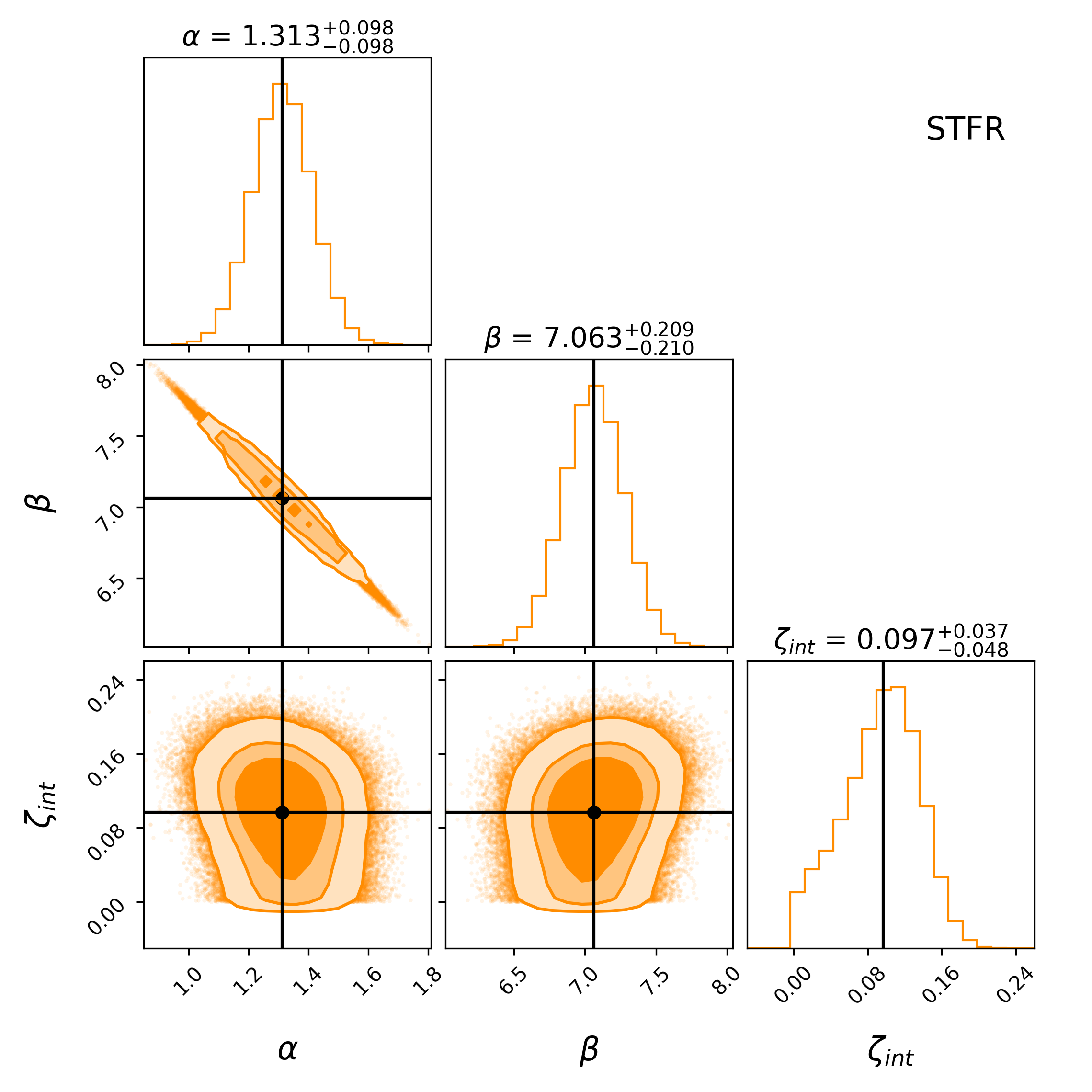}
		\includegraphics[width=\columnwidth]{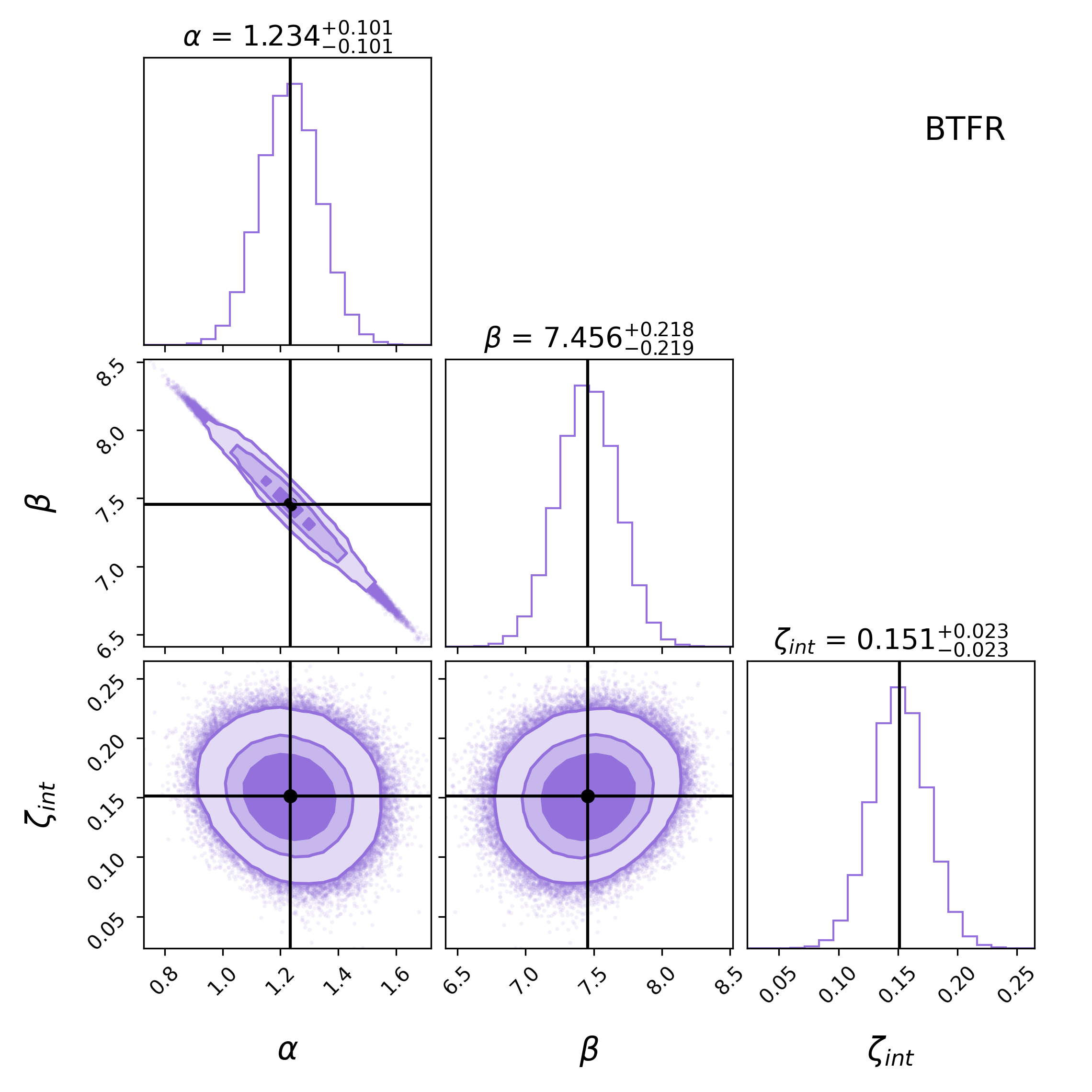}
		\caption{\textit{Upper Panel:} Stellar and Baryonic Tully-Fisher Relations (STFR and BTFR) obtained using a vertical likelihood method, presented in the left and right panels, respectively. The blue filled circles represent the data from \citet{GS23}, with gray error bars denoting uncertainties on each measurements. The solid orange and purple lines shows the best-fit curves obtained in this study, accompanied by the shaded regions representing the $3\sigma$ intrinsic scatter for the STFR and BTFR, respectively. The bottom right corner of each plot displays the best-fit parameters. Additionally, the blue lines correspond to comparisons with local studies, while red lines represent the high-redshift data, as indicated in the upper left legend of each plot. \textit{Lower Panel:} Posterior distributions (corner plots) resulting from the MCMC fitting process for the STFR and BTFR are shown in the left and right plots, respectively. The contours within these corner plots illustrate the 68\%, 90\%, and 99\% credible intervals.}
		\label{fig:TFR1_vert}
\end{figure*}

In this section we briefly discuss the results obtained using vertical likelihood that minimizes the intrinsic scatter in the vertical direction as defined in Eq. \eqref{eq:vert-ll1}. In Fig. \ref{fig:TFR1_vert}, we show the best-fit and corner plots for the STFR and BTFR on the full GS23 dataset in the left and right panels respectively. For STFR, we report $\alpha = 1.31\pm0.10$, $\beta=7.063\pm0.21$ and $\zeta_{int}=0.097$. Similarly for BTFR we find, $\alpha=1.23\pm0.1$, $\beta=7.46\pm0.22$  and $\zeta_{int}=0.15$. It is interesting to note that the slopes for the STFR and BTFR differ from the orthogonal best fit slopes by 1.72 dex and 1.98 dex respectively.

\end{appendix}

\end{document}